\documentclass[a4paper,man]{apa6}
\usepackage{IEEEtrantools}
\usepackage[english]{babel}
\usepackage[utf8x]{inputenc}
\usepackage{amsmath,amsthm,amssymb}
\usepackage{graphicx}
\usepackage{bm}
\usepackage[colorinlistoftodos]{todonotes}
\usepackage[authoryear]{natbib} 
\bibpunct{(}{)}{;}{a}{,}{,} 
\usepackage[flushleft]{threeparttable}
\usepackage{booktabs,caption}
\usepackage{colortbl}
\usepackage{graphicx}
\usepackage{pdflscape}
\usepackage{caption}
\usepackage{subcaption}
\usepackage{enumitem}
\usepackage{graphicx}
\usepackage{bm}
\usepackage{textcomp}
\usepackage{textgreek}
\usepackage{multicol}
\usepackage{setspace}
\usepackage{IEEEtrantools}
\usepackage{booktabs}
\usepackage{authblk}

\DeclareMathOperator{\Cov}{\mathbb{C}ov}

\title{An Instrumental Variable Estimator for Mixed Indicators: Analytic Derivatives and Alternative Parameterizations}
\shorttitle{Mixed Indicator Derivatives}
\author[1]{Zachary F. Fisher}
\author[1]{Kenneth A. Bollen}

\affil[1]{University of North Carolina at Chapel Hill}

\abstract{Methodological development of the Model-implied Instrumental Variable (MIIV) estimation framework has proved fruitful over the last three decades. Major milestones include Bollen's (1996) original development of the MIIV estimator and its robustness properties for continuous endogenous variable SEMs, the extension of the MIIV estimator to ordered categorical endogenous variables (Bollen \& Maydeu-Olivares, 2007), and the introduction of a Generalized Method of Moments (GMM) estimator (Bollen, Kolenikov \& Bauldry, 2014). This paper furthers these developments by making several unique contributions not present in the prior literature: (1) we use matrix calculus to derive the analytic derivatives of the PIV estimator, (2) we extend the PIV estimator to
apply to any mixture of binary, ordinal, and continuous variables, (3) we generalize the PIV model to include intercepts and means, (4) we devise a method to input known threshold values for ordinal observed variables, and (5) we enable a general parameterization that permits the estimation of means, variances, and covariances of the underlying variables to use as input into a SEM analysis with PIV. An empirical example illustrates a mixture of continuous variables and ordinal variables with fixed thresholds. We also include a simulation study to compare the performance of this novel estimator to WLSMV.}

\begin{document}
\makeatletter
\@ifundefined{@affil}{\def\@affil{~}{}}
\makeatother
\maketitle

\section{Introduction}

Classic Structural Equation Models (SEMs) treated all endogenous variables as continuous.  Contemporary SEM research has paid more attention to binary, ordinal, and other noncontinuous endogenous variables \citep{arminger1988, muthen1984, muthen1993, joreskog1994}.  Item response theory (IRT), of course, has a long history of treating measurement models with binary or ordinal measures \citep[e.g.][]{thurstone1925, lazarsfeld1950,bock1972,lord1968}.  A number of estimators have also been proposed for SEMs with categorical endogenous variables, including the Diagonally Weighted Least Squares (DWLS), Full-Information Maximum Likelihood (FIML), Pairwise Maximum Likelihood \citep[PML]{katsikatsou2012} and Polychoric Instrumental Variable \citep[PIV]{bollen2007} estimators. 

Recent research on the PIV estimator has revealed some promising features. \citet{nestler2013} found the PIV estimator to be as accurate as the Unweighted Least Squares (ULS) and DWLS estimators for correctly specified models and more robust in the presence of structural misspecifications.  Similarly, \citet{jin2016} found the PIV estimates to be as good as those from ULS and DWLS when the equation in question was correctly specified (and no additional misspecifications were present in the model), and less biased for correctly specified equations when specification errors were present in the model at large.  

These are important findings as it is far more likely our models are approximations rather than perfect realizations of the process under study. Furthermore, for system-wide estimators the bias induced by even minor misspecifications, such as an omitted factor loading,  have been found to linearly increase with each additional misspecification \citep{yang2010}. \citet{jin2018} proposed an equation-by-equation overidentification test for DWLS analogous to that used with Model Implied Instrumental Variables (MIIVs) 2SLS for continuous outcomes \citep{kirby2009}.

\textcolor{black}{Despite the PIV's valuable properties and extensions, it has several restrictions.  Bollen and Maydeu-Oliveres' (2007) current PIV estimator assumes that all observed variables are binary or ordinal.  They do not consider the common situation where a mix of binary, ordinal, censored, and continuous variables are analyzed.  This limits the applicability of their results.   Further, to simplify the situation they assumed that all the continuous variables that underlie the binary and ordinal variables had means of zero and variances of one.}  

\textcolor{black}{With continuous variables, researchers routinely estimate their means, variances, and covariances as part of the analyses. This practice is particularly valuable in longitudinal data where the differences in central tendency or variances are of research interest.  Similarly, this is true in multiple group analysis and invariance testing. A parameterization that allows estimated means, variances, and covariances creates greater continuity between analysis of continuous variables and those with ordinal variables rather than the dominant practice of assuming means of zero and variances of one for the underlying variable. In addition, many ordinal variables are categorized versions of continuous variables due to the constraints of survey research. For instance, a questionnaire item on household income collapsed into ranges such as “less than \$20,000, \$20,000 to \$40,000,..., greater than \$100,000” creates an ordinal variable from a continuous variable. Similarly, questionnaire responses on education, hours watching TV, number of alcohol drinks, cigarettes smoked, time on the internet, etc. often appear in ordinal categories that are collapsed versions of continuous variables. In such cases, the thresholds are known constants (e.g., \$20,000, \$40,000, etc. for income) that researchers could enter rather than estimate and thereby scale the underlying continuous variable in a metric close to that they would have obtained with the continuous variable. \citet{joreskog2002} proposed using thresholds as a way of estimating means and variables, but he restricted his alternative parameterization to setting the first two thresholds of an ordinal variable to 0 and 1. Our paper provides a more general approach to a mean, variance, and covariance parameterization that is applicable to DWLS as well as the MIIV approach.}
 
 \textcolor{black}{Another simplification Bollen and Maydeu-Oliveras (2007) use are numerical derivatives rather than analytic derivatives when developing estimates of the standard errors and significance tests.   The numerical derivatives are computationally slower and less accurate than analytic derivatives.  However, the analytic derivatives require matrix derivatives and are time consuming to find, despite their advantages.  One advantage is that once the analytic derivatives are available, they enable standard error estimates when a covariance (correlation) matrix and means are input for analysis as long as the asymptotic covariances of the covariance (correlation) matrix among the underlying variables and their means are available.} 

\textcolor{black}{Among the unique contributions of this paper are: (1) we use matrix calculus to derive the analytic derivatives of the PIV estimator, (2) we extend the PIV estimator to apply to any mixture of binary, ordinal, and continuous variables, (3) we generalize the PIV model to include intercepts and means so the analyst can include these as part of the modeling,  (4) we devise a method to input known threshold values for ordinal observed variables, and (5) we enable a general parameterization that permits the estimation of means, variances, and covariances of the underlying variables to use as input into a SEM analysis with PIV. An empirical example illustrates a mixture of continuous variables and ordinal variables with fixed thresholds.  We also include a simulation to compare the performance of this modified PIV estimator to WLSMV.}

\section*{The General Model and Assumptions}
We begin with a modification of a general structural equation model \footnote{Our notation is similar to the LISREL notation of Joreskog \& Sorbom (1978) except that we do not include exogenous latent variables and their corresponding indicators and we use $\mathbf{y}^{*}$ instead of y to refer to the underlying indicators of $\boldsymbol{\eta}$.} where the latent variable model is
\begin{equation}
\label{full1}
\boldsymbol{\eta} = 
  \boldsymbol{\alpha}_{\eta} +
  \mathbf{B} \boldsymbol{\eta} +
  \boldsymbol{\zeta},
\end{equation}
\noindent and $\boldsymbol{\eta}$ is a $m \times 1$ vector of latent endogenous variables, $\boldsymbol{\alpha}_{\eta}$ is a $m \times 1$ vector of constant intercept terms, and $\mathbf{B}$ is a $m \times m$ matrix of regression coefficients relating the endogenous variables. The equation disturbances are  contained in the $m \times 1$  vector $\boldsymbol\zeta$ and their variances and (cross-equation) covariances in  the $m \times m$ covariance matrix $\boldsymbol{\Sigma}_{\zeta}$. We assume $\mathbb{E}(\boldsymbol\zeta)=0$ and that $\boldsymbol\zeta$ is uncorrelated with any exogenous variables in $\boldsymbol\eta$. Furthermore, we propose the measurement model
\begin{IEEEeqnarray}{rCl}
\label{full2}
\mathbf{y}^{*} & = &
  \boldsymbol{\alpha}_{y} +
  \boldsymbol{\Lambda}_{y} \boldsymbol{\eta} +
  \boldsymbol{\varepsilon}
\end{IEEEeqnarray}
\noindent where $\mathbf{y}^{*}$ is a $p \times 1$ vector of continuous response variables, $\boldsymbol{\Lambda}_{y}$ is a $p \times m$ matrix of factor loadings, and $\boldsymbol{\varepsilon}$ is a  $p \times 1$ vector of errors or unique factors. We assume the errors have mean zero, $\mathbb{E}(\boldsymbol\varepsilon)=0$, and are uncorrelated with the latent variables,  $\mathbb{C}ov(\boldsymbol\varepsilon,\boldsymbol\eta)=0$. 

\textcolor{black}{An additional note describing the auxiliary measurement model linking $\mathbf{y}^{*}$ to the vector of observed variables $\mathbf{y}$ is warranted. To clarify our notation we note that $p$ is the total number of observed variables in the system, $r$ is the number of continuous observed variables, $s$ is the number of observed ordered categorical values, and $p=s+r$. If ${y}_{j},\dots,{y}_{r}$ represent continuous variables then $y_{j}=y^{*}_{j}$ for $j=1,\dots,r$. Conversely, if ${y}_{j}$ for $j=r+1,\dots,r+s$ represent binary or ordinal variables then $y_{j}^{*}$ is categorized according to the threshold parameters, $\boldsymbol{\tau}_{j}$, such that $y_{j} = c$ if $\tau_{j,c} < y_{j}^{*} < \tau_{j,c+1}$, where $c=0,...,C_{j}-1$. Here, $C_{j}$ is the total number of discrete values the ordered categorical variable $y_{j}$ can take and $C_{j}-1$ is the total number of thresholds to be estimated as $\tau_{j,0}=-\infty$ and $\tau_{j,C_{j}}=\infty$.}

\section*{A General Estimation Procedure}

\textcolor{black}{We partition our estimation procedure into three distinct steps.  In the first step we obtain consistent estimates of the first and second order sample statistics required for the subsequent analysis. In the second step, we develop alternative parameterizations for the $\mathbf{y}^{*}$ variables such that it becomes possible to construct an unconstrained covariance matrix $\boldsymbol{\Sigma}^{*}$ and mean vector $\boldsymbol{\mu}^{*}$ to serve as input to the analysis. Again using results from multivariate calculus, we obtain closed-form solutions for the limiting covariance matrix of $\upsilon({\hat{\boldsymbol{\Sigma}}^{*}})-\upsilon({\boldsymbol{\Sigma}^{*}})$ and $\upsilon({\hat{\boldsymbol{\mu}}^{*}})-\upsilon({\boldsymbol{\mu}^{*}})$. Here, the $\upsilon(\cdot)$ operator stacks the unique elements of a patterned matrix column-wise as in \citet{magnus1983}. Finally, in step 3 we derive the PIV coefficients corresponding to the full model parameters and analytic derivatives required to approximate their large sample variances. Table~\ref{algo} provides a summary of this algorithm.}\\
\subsection{First and Second Order Sample Statistics}
Here we obtain the first and second order sample statistics for the subsequent analysis. In the case of the model described by \citet{bollen2007} these statistics amount to the thresholds and polychoric correlations among the ordered categorical variables. With continuous variables included in $\mathbf{y}$ we must also estimate the means and covariance elements. Finally, we must also estimate the polyserial correlations among the continuous and categorical variables. 

\textcolor{black}{A number of methods have been described for obtaining these quantities \citep{olsson1979, muthen1984, joreskog1994, lee1995, katsikatsou2012, monroe2018}, and theoretically we could use any of these approaches.  However, for didactic purposes the results from  \citet{olsson1979} and \citet{muthen1984}  suffice.  We obtain the means and thresholds from the univariate margins. Let us consider a single variable $y_{j}$ from the set of $1,\dots,p$ observed variables. If $y_{j}$ is binary or ordinal, the most common convention to identify the thresholds  is to set $\mu_{j}=0$ and $\sigma_{jj}=1$. This enables us to estimate $\hat{\boldsymbol{\tau}}_{j}$  from the univariate margins using maximum likelihood.  If $y_{j}$ is a continuous variable,  $\mu_{j}$ and $\sigma_{jj}$ are freely estimated.  Subsequently, for every pairwise combination of ordered categorical variables the thresholds are held fixed at the estimates obtained from the univariate margins and the correlation structure ($\hat{\sigma}_{jk}$) for $j \neq k$ is obtained.  As a result of this conditional estimation procedure $\hat{\sigma}_{jk}$ is considered a pseudo-maximum likelihood estimate.} 

Now, collecting these sample statistics we have $\hat{\boldsymbol{\omega}} = (\hat{\boldsymbol{\mu}}_{y^{*}}, \hat{\boldsymbol{\tau}},\upsilon(\hat{\boldsymbol{\Sigma}}_{y^{*}}))$, where $\boldsymbol{\Sigma}_{y^{*}}$ takes the following form

\newpage

\begin{IEEEeqnarray}{rCl}
\label{sigystar}
\boldsymbol{\Sigma}_{y^{*}}=\begin{bmatrix} 
    		1 &  &   & &&\\
    		\vdots & \ddots & & &&\\
    		\rho_{y^{*}_{s,1}} & \dots & 1 & &\\
		\tilde{\rho}_{y^{*}_{s+1,1}} &\dots &\tilde{\rho}_{y^{*}_{s+1,s}}& \sigma_{y^{*}_{s+1,s+1}} &\\
		\vdots & \ddots &\vdots&\vdots& \ddots \\
		\tilde{\rho}_{y^{*}_{s+r,1}} & \dots  &\tilde{\rho}_{y^{*}_{s+r,s}} & \sigma_{y^{*}_{s+r,s+1}}
			&\dots&  \sigma_{y^{*}_{s+r,s+r}}\\
\end{bmatrix}
\end{IEEEeqnarray}
where $\rho_{y^{*}_{j,k}} =  \mathbb{C}or(y^{*}_{j},y^{*}_{k}) \: \forall \:j \neq k, ; \: j,k< s $ represent the polychoric correlations among the $s$ ordered categorical variables, 
$\tilde{\rho}_{y^{*}_{j,k}} =  \mathbb{C}or(y^{*}_{j},y^{*}_{k}) \: \forall \: s < j \leq s+r, \: k< s$ represent the polyserial correlations among the $s$ ordered categorical variables and $r$ continuous variables, and 
$\sigma_{y^{*}_{j,k}} =  \mathbb{C}ov(y^{*}_{j},y^{*}_{k}) \: \forall \: s < (j,k) \leq s + r$ represent the covariances among the $r$ continuous variables. The mean vector is $\boldsymbol{\mu}^{'}_{{y}^{*}} = (0_{1},\dots,0_{s},\mu_{y^{*}_{s+1}}, \dots,\mu_{y^{*}_{s+r}})$.
For the preceding procedure the consistency of $\hat{\boldsymbol{\omega}}$, $\mathrm{plim}\:\hat{\boldsymbol{\omega}}= {\boldsymbol{\omega}}$, and the assumptions that underlie the proof are given by  \citet[Appendix A]{muthen1995}. Furthermore,  \citet[pp. 494-498]{muthen1995}  provide a convenient procedure for constructing $\hat{\boldsymbol{\omega}}$ and the conditions under which it achieves asymptotic normality,
\begin{IEEEeqnarray}{rCl}
\label{normsigma}
n^{1/2}(\hat{\boldsymbol{\omega}}-{\boldsymbol{\omega}})\xrightarrow[]{d} \mathcal{N}(\mathbf{0}, \boldsymbol{\Sigma}_{\omega}).
\end{IEEEeqnarray}
\noindent \textcolor{black}{There are a number of ways for estimating these quantities and their asymptotic covariance matrix, including methods proposed by \citet{olsson1979}, \citet{muthen1984}, \citet{joreskog1994}, \citet{lee1995}, \citet{katsikatsou2012}, and \citet{monroe2018}. The analytic derivatives provided here are compatible with any of these methods.} 
\subsection{An Unconstrained Covariance Matrix and Mean Vector}
When the means and variances of the $\mathbf{y}^{*}$ variables are of interest, it is possible to estimate these quantities rather than constrain them to $0$ and $1$, respectively. For this reason we also consider alternative parameterizations for ordered categorical variables with three or more response options.  \citet{joreskog2002} showed a specific alternative parameterization where the first two thresholds of $y^{*}_{j}$ are fixed to zero and one, respectively.  From these identifying constraints the mean and variance of $y^{*}_{j}$ can be estimated.  We generalize this idea for additional parameterizations in the following manner. Consider the latent response variable $y^{*}_{j}$ underlying the ordered categorical variable $y_{j}$. In step 1 we assume $y^{*}_{j} \sim \mathcal{N}(0,1)$ , where $y^{*}_{j}$ is determined only up to a monotonic transformation of $y_{j}$.  To retain the normality assumption we can allow any arbitrary linear transformation of $y_{j}^{*}$. If we only wish to estimate the mean (while keeping $\mathbb{V}ar(y^{*}_{j})$ fixed to one) we shift the distribution of $y^{*}_{j}$ by a constant, $q_{1}$, such that the transformed distribution $\ddot{y}^{*}_{j} \sim \mathcal{N}(q_{1},1)$.  If in addition to the mean we also wish to estimate the variance of $y^{*}_{j}$, we shift  and scale the distribution of $y_{j}^{*}$ such that $\ddot{y}^{*}_{j} \sim \mathcal{N}(q_{1}q_{2},q_{2}^{2})$.

To provide a concrete example of this transformation let $\tau_{j,a}$ and $\tau_{j,b}$ be the first two thresholds of the latent response variable $y^{*}_{j}$ underlying the four-category observed variable $y_{j}$. In Step 1 of the analysis, we estimated all three thresholds of $y^{*}_{j}$ assuming $\mu_{j}=0$ and $\sigma_{jj}=1$. Now, let $\ddot{\tau}^{*}_{j,a}$ and $\ddot{\tau}_{j,b}$ be the corresponding first two thresholds under an alternative parameterization where we wish to instead estimate $\mu_{j}$ and $\sigma_{jj}$. Accordingly we must now fix any two of the previously estimated thresholds to some constant values (e.g. $\ddot{\tau}_{j,a}=0$ and $\ddot{\tau}_{j,b}=1$). Using the following algebraic results 
\begin{IEEEeqnarray}{rCl}
\label{C1C2}
q_{1} & = & -\frac{\tau^{}_{j,a}\ddot{\tau}_{j,b}-\tau^{}_{j,b}\ddot{\tau}_{j,a}}{\ddot{\tau}_{j,b}-\ddot{\tau}_{j,a}} \\
q_{2} & = & \ddot{\tau}_{j,a}(q_{1}+\tau_{j,a})^{-1}
\end{IEEEeqnarray}
it becomes possible to estimate the mean, variance, as well as the remaining thresholds, $\ddot{\tau}_{j,k} = (\tau_{j,k}+q_{1})q_{2}$, for the transformed distribution. 

To construct $\boldsymbol{\Sigma}_{y^{*}}$ and $\boldsymbol{\mu}_{y^{*}}$ under the alternative parameterization we must reexpress the scalar representation given above in matrix form. To do this we provide the most general form, the case where one wishes to estimate both the mean and variance for each of the $s$ ordered categorical variables by fixing two of the thresholds to constant values. For latent response variable $y_{j}^{*}$ we denote the two fixed value thresholds as $\ddot{\tau}_{j,a}$ and $\ddot{\tau}_{j,b}$ and their freely estimated counterparts under the standard parameterization as $\tau_{j,a}$ and $\tau_{j,b}$. Building on this notation we let $\mathbf{D}_{a}=\mathrm{diag}(\tau_{1,a},\dots,\tau_{s,a},1_{s+1},\dots,1_{s+r})$, $\mathbf{D}_{b}=\mathrm{diag}(\tau_{1,b},\dots,\tau_{s,b},0_{s+1},\dots,0_{s+r})$,  $\ddot{\mathbf{D}}_{a}=\mathrm{diag}(\ddot{\tau}_{1,a},\dots,\ddot{\tau}_{s,a},1_{s+1},\dots,1_{s+r})$, and $\ddot{\mathbf{D}}_{b}=\mathrm{diag}(\ddot{\tau}_{1,b},\dots,\ddot{\tau}_{s,b},0_{s+1},\dots,0_{s+r})$. Lastly we let $\mathbf{D}_{\tau,k}=\mathrm{diag}(\tau_{1,k},\dots,\tau_{s,k},0_{s+1},\dots,0_{s+r})$. If $\tau_{j,k}$ does not exist for $y^{*}_{j}$, the $j$the diagonal of $\mathbf{D}_{\tau, k}$ equals zero.
Then, in matrix form we can express the scalar constants $q_{1}$ and $q_{2}$ from above as
\begin{IEEEeqnarray}{rCl}
\label{C1C2}
\mathbf{Q}_{1} & = & - \mathbf{D}_{a} \ddot{\mathbf{D}}_{b} - 
	 \mathbf{D}_{b} \ddot{\mathbf{D}}_{a} (\ddot{\mathbf{D}}_{b}-\ddot{\mathbf{D}}_{a})^{-1} \\
\mathbf{Q}_{2} & = &\ddot{\mathbf{D}}_{a} (\mathbf{Q}_{1}+ \mathbf{D}_{a})^{-1}
\end{IEEEeqnarray}
along with the thresholds, mean vector and covariance matrix
\begin{IEEEeqnarray}{rCl}
\label{taudot}
\ddot{\boldsymbol{\tau}}_{y^{*},k} & = &\mathrm{diag}[(\mathbf{D}_{\tau,k} +   \mathbf{Q}_{1})\mathbf{Q}_{2}],\\
\label{mudot}
\ddot{\boldsymbol{\mu}}_{y^{*}} & = & \mathrm{diag}(\mathbf{Q}_{1}\mathbf{Q}_{2})+\boldsymbol{\mu}_{y^{*}}, \\
\label{sigdot}
\ddot{\boldsymbol{\Sigma}}_{y^{*}} & = &\mathbf{Q}_{2}\boldsymbol{\Sigma}_{y^{*}}\mathbf{Q}_{2}.
\end{IEEEeqnarray}
\subsubsection{Approximate Asymptotic Distribution of $\hat{\boldsymbol{\pi}}$}
\textcolor{black}{Let $\boldsymbol{\pi}$ contain the free elements of the unconstrained covariance matrix and mean vector.} To obtain the asymptotic distribution of $\boldsymbol{\pi}$ we employ the multivariate-delta method \citep{cramer1999}.  The multivariate-delta method provides a convenient approach for resolving the large sample distribution of a vector function of a multinormally distributed random vector.  Earlier it was shown that 
$n^{1/2}(\hat{\boldsymbol{\omega}}-{\boldsymbol{\omega}})\xrightarrow[]{d} \mathcal{N}(\mathbf{0}, \boldsymbol{\Sigma}_{\omega})$.  As the elements of $\boldsymbol{\pi}$ are a function of $\boldsymbol{\omega}$ we can express the large sample variances of $\boldsymbol{\pi}$ as
\begin{IEEEeqnarray}{rClCl} 
\label{varpi}
\mathbb{V}ar(\boldsymbol{\pi}) &=& \left.\frac{\partial\:\boldsymbol{\pi}(\hat{\boldsymbol{\omega}})} 
{\partial\: \boldsymbol{\omega}}\right|^{'}_{ \boldsymbol{\omega}= \hat{\boldsymbol{\omega}}}
\hat{\boldsymbol{\Sigma}}_{\omega}
 \left.\frac{\partial\:\boldsymbol{\pi}(\hat{\boldsymbol{\omega}})} 
{\partial\: \boldsymbol{\omega}}\right|_{ \boldsymbol{\omega}= \hat{\boldsymbol{\omega}}}.
\end{IEEEeqnarray}
where $\partial\:\boldsymbol{\pi}(\hat{\boldsymbol{\omega}})/  \partial\: \boldsymbol{\omega}$ is the Jacobian matrix containing the first order partial derivatives of $\ddot{\boldsymbol{\Sigma}}_{y^{*}}$, $\ddot{\boldsymbol{\mu}}_{y^{*}}$ and $\ddot{\boldsymbol{\tau}}_{y^{*}}$ evaluated with respect to  $\mathbf{D}_{a}$, $\mathbf{D}_{b}$, $\mathbf{D}_{\tau,k}$, $\boldsymbol{\Sigma}_{y^{*}}$ and $\boldsymbol{\mu}_{y^{*}}$.
The Jacobian matrix is comprised of the partial derivatives of $\ddot{\boldsymbol{\tau}}_{y^{*}}$,$\ddot{\boldsymbol{\mu}}_{y^{*}}$ and $\ddot{\boldsymbol{\Sigma}}_{y^{*}}$  with respect to the freely varying elements in the $p \times 1$ vector $\boldsymbol{\mu}$, the $p \times p$ diagonal matrices $\mathbf{D}_{a}$, $\mathbf{D}_{b}$ and $\mathbf{D}_{\tau,k}$ and the $p \times p$ correlation/covariance matrix $\boldsymbol{\Sigma}_{y^{*}}$, 
\begin{IEEEeqnarray}{rClCl} 
\label{pd_Sigma_Da}
\frac{\partial\: \upsilon(\ddot{\boldsymbol{\Sigma}}_{y^{*}})} 
       {\partial\: \upsilon(\mathbf{D}_{a})^{'}} 
 &=& 
\boldsymbol{\Delta}^{+}      
	\big[
	(\mathbf{Q}_{2}\boldsymbol{\Sigma}_{y^{*}} \otimes \mathbf{I} +
	\mathbf{I} \otimes \mathbf{Q}_{2}\boldsymbol{\Sigma}_{y^{*}}) 
	\frac{\partial\: \mathbf{Q}_{2}} 
       {\partial\: \mathbf{D}_{a}^{'}} 
	\big]
\boldsymbol{\Delta}, \\
\label{pd_Sigma_Db}
\frac{\partial\: \upsilon(\ddot{\boldsymbol{\Sigma}}_{y^{*}})} 
       {\partial\: \upsilon(\mathbf{D}_{b})^{'}} 
 &=& 
\boldsymbol{\Delta}^{+}      
	\big[
	(\mathbf{Q}_{2}\boldsymbol{\Sigma}_{y^{*}} \otimes \mathbf{I} +
	\mathbf{I} \otimes \mathbf{Q}_{2}\boldsymbol{\Sigma}_{y^{*}}) 
	\frac{\partial\: \mathbf{Q}_{2}} 
       {\partial\: \mathbf{D}_{b}^{'}} 
	\big]
\boldsymbol{\Delta}, \\
\label{pd_Sigma_Sig}
\frac{\partial\: \upsilon(\ddot{\boldsymbol{\Sigma}}_{y^{*}})} 
       {\partial\: \upsilon(\boldsymbol{\Sigma}_{y^{*}})^{'}} 
 &=& 
\boldsymbol{\Delta}^{+}      
(\mathbf{Q}_{2} \otimes \mathbf{Q}_{2})
\boldsymbol{\Delta}, \\
\label{pd_Mu_Da}
\frac{\partial\: \upsilon(\ddot{\boldsymbol{\mu}}_{y^{*}})} 
       {\partial\: \upsilon(\mathbf{D}_{a})^{'}} 
 &=& 
\boldsymbol{\Delta}^{+}      
	\big[   
	(\mathbf{Q}_{2} \otimes \mathbf{I})	
	\frac{\partial\: \mathbf{Q}_{1}} {\partial\: \mathbf{D}_{a}^{'}} +
	(\mathbf{I}\otimes \mathbf{Q}_{1} )	
	\frac{\partial\: \mathbf{Q}_{2}} {\partial\: \mathbf{D}_{a}^{'}} 	
	\big]
\boldsymbol{\Delta}, \\
\label{pd_Mu_Db}
\frac{\partial\: \upsilon(\ddot{\boldsymbol{\mu}}_{y^{*}})} 
       {\partial\: \upsilon(\mathbf{D}_{b})^{'}} 
 &=& 
\boldsymbol{\Delta}^{+}      
	\big[   
	(\mathbf{Q}_{2} \otimes \mathbf{I})	
	\frac{\partial\: \mathbf{Q}_{1}} {\partial\: \mathbf{D}_{b}^{'}} +
	(\mathbf{I}\otimes \mathbf{Q}_{1} )	
	\frac{\partial\: \mathbf{Q}_{2}} {\partial\: \mathbf{D}_{b}^{'}} 	
	\big]
\boldsymbol{\Delta}, \\
\label{pd_Mu_Mu}
\frac{\partial\: \upsilon(\ddot{\boldsymbol{\mu}}_{y^{*}})} 
       {\partial\: \upsilon(\boldsymbol{\mu}_{y^{*}})^{'}} 
 &=& 
\boldsymbol{\Delta}^{+} \boldsymbol{\Delta}, \\
\label{pd_Tau_Tau}
\frac{\partial\: \upsilon(\ddot{\boldsymbol{\tau}})} 
       {\partial\: \upsilon(\mathbf{D}_{\tau,k})^{'}} 
 &=& 
\boldsymbol{\Delta}^{+} 
\mathbf{Q}_{2}
\boldsymbol{\Delta}, \\
\label{pd_Tau_Da}
\frac{\partial\: \upsilon(\ddot{\boldsymbol{\tau}})} 
       {\partial\: \upsilon(\mathbf{D}_{a})^{'}} 
 &=& 
\boldsymbol{\Delta}^{+}      
	\big[   
	(\mathbf{Q}_{2} \otimes \mathbf{I})	
	\frac{\partial\: \mathbf{Q}_{1}} {\partial\: \mathbf{D}_{a}^{'}} +
	(\mathbf{I}\otimes (\mathbf{Q}_{1} + \mathbf{D}_{\tau}))	
	\frac{\partial\: \mathbf{Q}_{2}} {\partial\: \mathbf{D}_{a}^{'}} 	
	\big]
\boldsymbol{\Delta}, \\
\label{pd_Tau_Db}
\frac{\partial\: \upsilon(\ddot{\boldsymbol{\tau}})} 
       {\partial\: \upsilon(\mathbf{D}_{b})^{'}} 
 &=& 
\boldsymbol{\Delta}^{+}      
	\big[   
	(\mathbf{Q}_{2} \otimes \mathbf{I})	
	\frac{\partial\: \mathbf{Q}_{1}} {\partial\: \mathbf{D}_{b}^{'}} +
	(\mathbf{I}\otimes (\mathbf{Q}_{1} + \mathbf{D}_{\tau}))	
	\frac{\partial\: \mathbf{Q}_{2}} {\partial\: \mathbf{D}_{b}^{'}} 	
	\big]
\boldsymbol{\Delta} ],
\end{IEEEeqnarray}
where
\begin{IEEEeqnarray}{rClCl} 
\label{pd_C2_Da}
	\frac{\partial\: \mathbf{Q}_{2}} 
       {\partial\: \mathbf{D}_{a}^{'}} 
 &=& - \big[
	(\mathbf{Q}_{1}+\mathbf{D}_{a})^{-1} \otimes  \ddot{\mathbf{D}}_{a}(\mathbf{Q}_{1}+\mathbf{D}_{a})^{-1}
\big]
(	\frac{\partial\: \mathbf{Q}_{1}} 
       {\partial\: \mathbf{D}_{a}^{'}} + \mathbf{I}), \\
\label{pd_C1_Da}
	\frac{\partial\: \mathbf{Q}_{1}} 
       {\partial\: \mathbf{D}_{a}^{'}} 
 &=& -
[ (\ddot{\mathbf{D}}_{b} - \ddot{\mathbf{D}}_{a})^{-1} \ddot{\mathbf{D}}_{b} \otimes \mathbf{I}], \\
\label{pd_C2_Db}
	\frac{\partial\: \mathbf{Q}_{2}} 
       {\partial\: \mathbf{D}_{b}^{'}} 
 &=& - \big[
	(\mathbf{Q}_{1}+\mathbf{D}_{a})^{-1} \otimes  \ddot{\mathbf{D}}_{a}(\mathbf{Q}_{1}+\mathbf{D}_{a})^{-1}
\big]
	\frac{\partial\: \mathbf{Q}_{1}} 
       {\partial\: \mathbf{D}_{b}^{'}}, \\
\label{pd_C1_Db}
	\frac{\partial\: \mathbf{Q}_{1}} 
       {\partial\: \mathbf{D}_{b}^{'}} 
 &=& -
[ (\ddot{\mathbf{D}}_{b} - \ddot{\mathbf{D}}_{a})^{-1} \ddot{\mathbf{D}}_{a} \otimes \mathbf{I}].
 \end{IEEEeqnarray}
\noindent and $\boldsymbol{\Delta}$ is a generalized duplication matrix, $\boldsymbol{\Delta}^{+}$ is a generalized elimination matrix \citep{magnus1983}.  Descriptions of these operators and their relation to patterned matrix derivatives are given in Appendix A. Now, let $L_{1}=[\frac{\partial\: \upsilon(\ddot{\boldsymbol{\Sigma}}_{y^{*}})}{\partial\: \upsilon(\mathbf{D}_{a})^{'}},\frac{\partial\: \upsilon(\ddot{\boldsymbol{\Sigma}}_{y^{*}})} 
{\partial\: \upsilon(\mathbf{D}_{b})^{'}} , \frac{\partial\: \upsilon(\ddot{\boldsymbol{\Sigma}}_{y^{*}})} 
{\partial\: \upsilon(\boldsymbol{\Sigma}_{y^{*}})^{'}} ]$, $L_{2}=[\frac{\partial\: \upsilon(\ddot{\boldsymbol{\mu}}_{y^{*}})} {\partial\: \upsilon(\mathbf{D}_{a})^{'}}, \frac{\partial\: \upsilon(\ddot{\boldsymbol{\mu}}_{y^{*}})} {\partial\: \upsilon(\mathbf{D}_{b})^{'}}, \frac{\partial\: \upsilon(\ddot{\boldsymbol{\mu}}_{y^{*}})} {\partial\: \upsilon(\boldsymbol{\mu}_{y^{*}})^{'}} ]$, and $L_{3}=[\frac{\partial\: \upsilon(\ddot{\boldsymbol{\tau}})}{\partial\: \upsilon(\mathbf{D}_{\tau})^{'}}, \frac{\partial\: \upsilon(\ddot{\boldsymbol{\tau}})} {\partial\: \upsilon(\mathbf{D}_{a})^{'}}, \frac{\partial\: \upsilon(\ddot{\boldsymbol{\tau}})} {\partial\: \upsilon(\mathbf{D}_{b})^{'}} ]$. Here $\frac{\partial\: \upsilon(\ddot{\boldsymbol{\tau}})}{\partial\: \upsilon(\mathbf{D}_{\tau})^{'}}$ is a block diagonal matrix composed of the $k=1,\dots,g$ jacobian matrices  $\frac{\partial\: \upsilon(\ddot{\boldsymbol{\tau}})}{\partial\: \upsilon(\mathbf{D}_{\tau,k})^{'}}$ and $g$ is the maximum number of free thresholds for the variables in $\mathbf{y}^{*}$. Finally, 
\begin{IEEEeqnarray}{rClCl} 
\label{L}
\frac{\partial\:\boldsymbol{\pi}(\hat{\boldsymbol{\omega}})} 
{\partial\: \boldsymbol{\omega}} &=&
[\mathbf{L}_{1}^{'},\mathbf{L}_{2}^{'},\mathbf{L}_{3}^{'}]^{'}
\end{IEEEeqnarray}
and the entries of $\partial\:\boldsymbol{\pi}(\hat{\boldsymbol{\omega}})/  \partial\: \boldsymbol{\omega}$ have been reordered to match the ordering of elements in  $\boldsymbol{\Sigma}_{\omega}$. Combining \eqref{varpi} and the properties of the multivariate-delta method, $n^{1/2}(\upsilon(\hat{\boldsymbol{\pi}})-\upsilon({\boldsymbol{\pi}}))\xrightarrow[]{d} \mathcal{N}(\mathbf{0}, \mathbb{V}ar(\boldsymbol{\pi}))$. With a consistent estimate of $\mathbb{V}ar(\boldsymbol{\pi})$ in hand we move to the estimation of the full model parameters.

\section{PIV Estimation}

In Stage 3 the correlation or covariance matrix (standard parameterization) or the unconstrained covariance matrix (alternative parameterization) from Step 2 is used to obtain point estimates for the parameters detailed in  \eqref{full1} and \eqref{full2}.  Furthermore, the large sample distribution of these parameters is derived. Although it is beyond the scope of this paper to present an exhaustive description of MIIV estimation to contextualize our procedure we provide the prerequisite details and point readers to the source material for more detailed exposition. 

\subsubsection{MIIV Estimation} The MIIV family of estimators \citep{bollen1996,bollen2007,bollen2014} begin by transforming the latent variable model into a system of estimating equations.  A detailed description of this transformation is given by \citet{bollen1996}.  A consequence of this transformation is that each equation's error will be a composite of errors from related equations. In most cases terms in this composite error will correlate with the endogenous right hand side variables. To address this endogeneity instrumental variables implied by the model specification itself (MIIVs) are ascertained for each equation in the system. To be useful these instruments must be related to endogenous variables and unrelated to the composite error, and a number of tests are available for assessing the validity of instrumental variables \citep{kirby2009}. For the purpose of this paper we focus on the MIIV-2SLS estimation itself, concentrating on the stages occurring after the model has been transformed into a system of estimating equations, however a brief description of the estimating equations is warranted. \\

\textcolor{black}{We begin by  scaling each latent variable in the model by choosing one of its indicators and setting its intercept to zero and its factor loading to identity. We label this variable the scaling indicator.} This choice allows us to reexpress each latent variable in terms of its scaling indicator minus its error. This is called the latent to observed variable transformation and proceeds by partitioning the $p$ manifest variables into two vectors, $\mathbf{y}^{(s)}$ and $\mathbf{y}^{(n)}$, where $\mathbf{y}^{(s)}$ contains the $m$ scaling indicators, and $\mathbf{y}^{(n)}$ contains the $p-m$ non-scaling indicators, such that $\mathbf{Y}^{'}=[\mathbf{y}^{(s)'},\mathbf{y}^{(n)'}]^{'}$. Using algebra we can then transform \eqref{full1} and \eqref{full2} into

\begin{IEEEeqnarray}{rCl} 
\label{full3}
	\mathbf{Y}
    & = & 
    \begin{bmatrix}  
  		\boldsymbol{\alpha}_{\eta} \\
        \boldsymbol{\alpha}_\mathrm{y}^{(n)} \\
	\end{bmatrix} +
    \begin{bmatrix}  
  		 \mathbf{B}  \\
         \boldsymbol{\Lambda}^{(n)} \\
	\end{bmatrix} \mathbf{y}^{(s)} +
     \begin{bmatrix}  
  		(\mathbf{I}-\mathbf{B}) & \mathbf{0} & \mathbf{I}\\
        - \boldsymbol{\Lambda}^{(n)} \:\: & \mathbf{I} & \mathbf{0} \\
	\end{bmatrix}
        \begin{bmatrix} 
         \boldsymbol{\varepsilon}^{(s)} \\
         \boldsymbol{\varepsilon}^{(n)} \\
         \boldsymbol{\zeta}  \\
	\end{bmatrix}
\end{IEEEeqnarray}
where $\boldsymbol{\alpha}_{\eta}$ is the $m \times 1$ vector of latent variable intercepts, $\boldsymbol{\alpha}_{y}^{(n)}$ is the $(p-m) \times 1$ vector of measurement model intercepts and $\boldsymbol{\Lambda}^{(n)}$ is the $(p-m) \times m$ factor loading matrix for the non-scaling indicators. \\

\textcolor{black}{The transformation above has translated the structural relations from the original model into a system of estimating equations.  Consolidating the composite disturbance from \eqref{full3} we can further simplify our notation to express this system as  $\mathbf{Y} = \mathbf{Z} \bm{\theta}_{B\Lambda} + {\mathbf{\widetilde{u}}}$ where $\mathbf{Z}$ is block-diagonal and contains all relevant regressors from $\mathbf{y}^{(s)}$, $\boldsymbol{\theta}_{B\Lambda}$ contains the free parameters in $\mathbf{B}$ and $\boldsymbol\Lambda$ from  \eqref{full1} and \eqref{full2} and ${\mathbf{\widetilde{u}}}$ contains the composite error terms for each equation. The difficulty in estimating $\boldsymbol{\theta}_{B\Lambda} $ from $\mathbf{y} = \mathbf{Z}\bm{\theta}_{B\Lambda}  + {\mathbf{\widetilde{u}}}$ results from the composite disturbance term ${\mathbf{\widetilde{u}}}$ which by construction will have a nonzero correlation with variables in ${\mathbf{Z}}$. To overcome this difficulty we can use the MIIV-2SLS estimator to obtain consistent estimates of $\bm{\theta}_{B\Lambda}$ where instruments for each equation are implied by the larger model specification. Let ${\mathbf{V}}$ be the block diagonal matrix containing these equation-specific instrumental variables.  For equation $j$, the matrix of IVs, ${\mathbf{V}}_{j}$, are valid if the following conditions are satisfied: (a) $\Cov({\mathbf{\widetilde{u}}}_{j},{\mathbf{V}}_{j}) = 0$ , (b) rank of $\Cov(\mathbf{V}_{j},\mathbf{Z}_{j})$ must be equal to the number of columns in  ${\mathbf{Z}}_{j}$, (c) $\Cov({\mathbf{V}}_{j})$ is nonsingular, and (d) $\Cov({\mathbf{Z}}_{j},{\mathbf{V}}_{j}) \neq 0$.}

\textcolor{black}{It is helpful to discuss these assumptions further and specifically note their implications for the SEM context. Assumption (a) is essential to ensure the consistency of the estimator. It requires the instrumental variables to be uncorrelated with the equation disturbance. We note that the covariances between the composite error and MIIVs can be read directly from the model-implied moments, and thus a correctly specified model will immediately point to the observed variables that satisfy (a). Condition (b) is used to ensure the parameters in equation $j$ are identified. This rank condition fails if the number of instruments is less than the number of explanatory variables in an equation, and an implication of this condition is that we must have at least one instrument for each explanatory variable. Condition (c) ensures the absence of perfect multicollinearity among the MIIVs. Finally, condition (d) is designed to ensure a sufficient association between the endogenous variables and MIIVs specified for equation $j$, marginal association of this condition leads to the \emph{weak instrument} problem.  In the context of SEM the weak instrument condition has received less attention.  One reason is because nonscaling indicators of the same latent variables are commonly MIIVs, this often leads to MIIVs that are moderate to strong.  This departs from many typical econometric cases where the instrument are auxiliary variables not part of the original model. However, the consequence of weak instruments are serious enough to warrant additional discussion.}

\textcolor{black}{In the presence of any nonzero correlation between the instruments and equation error, and a weak correlation between the endogenous regressors and instruments, or weak instruments, instrumental variable estimators can become biased and the rejection rate of the null hypothesis of no effect too large. \citet{phillips1989} was one of the first to draw attention to the consequences of weak instruments, however, the developments of \citet{stock2005} have become the \emph{de facto} framework for testing weak instruments in statistical software. The \citet{stock2005} approach based their approach on quantifying the relationship between instrument strength and the resulting bias of the instrumental variable estimator relative to the bias of OLS (and the size of inferential tests).  In the case of a single endogenous regressor \citet{stock2005} employ a $F$ statistic obtained from a regression of the endogenous regressor on the instrumental variables. In the case of multiple endogenous regressors the minimum-eigenvalue statistic of \citet{cragg1993} is used. Importantly, \citet{stock2005} provide tables for the critical values in both cases.}

\textcolor{black}{Unfortunately, the inferential methods developed by \citet{stock2005} rely heavily on the assumption of homoskedasticity and are intended for continuous variables only. Another widely used measure of instrument strength is Shea's (\citeyear{shea1997}) partial $R^2$.  Shea's partial $R^2$ is useful in the current context as it (a) has a straightforward interpretation, (b) can handle multiple endogenous variables in a single equation, and (c) its use does not rely on identifying the appropriate sampling distribution of the statistic.  For these reasons we present a version of the statistic adapted to the use of covariance matrices as input, $R^{2}_{S} = \mathbf{S}_{VX}^{'}\mathbf{S}_{VV}^{-1}\mathbf{S}_{VX}^{'}\mathbf{S}_{XX}^{-1}$ where $\mathbf{S}_{VX}$ is the covariance matrix of the MIIVs and endogenous regressors, while $\mathbf{S}_{VV}$ and $\mathbf{S}_{XX}$ are the covariance matrices of the MIIVs and endogenous variables, respectively. Although this statistic is calculated at the equation level we have omitted the subscript $j$ for clarity.}

\subsubsection{Estimation of PIV Regression Model Parameters}
Let $\boldsymbol{\theta}$ contain the full set of model parameters  from \eqref{full1} and \eqref{full2}. It is convenient to separate $\boldsymbol{\theta}$ into $\boldsymbol{\theta}_{1}$ and $\boldsymbol{\theta}_{2}$, where each contains the mathematically independent elements of ($\boldsymbol{\alpha}_{\eta}$, $\boldsymbol{\alpha}_{y^{*}}$, $\boldsymbol{\Lambda}$, $\mathbf{B}$) and ($\boldsymbol{\Sigma}_{\zeta}$, $\boldsymbol{\Sigma}_{\varepsilon}$), respectively. Furthermore, we partition $\boldsymbol{\theta}_{1}$ into $\boldsymbol{\theta}_{B\Lambda}=[\upsilon( \boldsymbol{\Lambda}), \upsilon(\mathbf{B})]$ and $\boldsymbol{\theta}_{\alpha}=[\upsilon(\boldsymbol{\alpha}_{\eta}), \upsilon(\boldsymbol{\alpha}_{y^{*}})]$ such that $\boldsymbol{\theta}_{1}=[\boldsymbol{\theta}_{\alpha},\boldsymbol{\theta}_{B\Lambda}]$.

To estimate the parameters in  $\boldsymbol{\theta}_{1}$ we employ a generalized method of moments (GMM) approach. GMM estimation is based on the idea that a set of parameters $\boldsymbol{\theta}$ from an overidentified model are related to a set of data $\mathbf{X}$ through a series of orthogonality conditions on the population moments, $\mathbb{E}[\mathbf{g}(\mathbf{x};\boldsymbol{\theta})]=0$ \citep{hansen1982}. Here we can define the vector $\mathbf{g}()$ as 
\begin{IEEEeqnarray}{rCl} 
\label{pivtheta1a}
\mathbf{g} = \mathbf{g}(\mathbf{x};\boldsymbol{\theta}) = \mathbf{V}(\mathbf{Y} - \mathbf{Z} \boldsymbol{\theta})
\end{IEEEeqnarray}
\noindent where $\mathbf{V}$  is a block-diagonal matrix of instrumental variables, $\mathbf{Y}$ is the vector of dependent variables, and $\mathbf{Z}$ is a block diagonal matrix of explanatory variables. As the input to our analysis is the joint unconditional covariance matrix $\boldsymbol{\Sigma}^{*}$, we can express the quantities $\mathbf{S}^{*}_{vy^{*}}$ and $\mathbf{S}^{*}_{vz}$ as the sample moments corresponding to $\boldsymbol{\Sigma}^{*}_{vy^{*}}$ and  $\boldsymbol{\Sigma}^{*}_{vz}$, respectively.  The goal in estimation is to choose $\boldsymbol{\widehat\theta}$ that minimizes the joint distance between the sample orthogonality conditions using the symmetric, positive-definite weight matrix $\mathbf{W}$, $N(\mathbf{S}^{*}_{vy^{*}}-\mathbf{S}^{*}_{vz}\widehat{\boldsymbol{\theta}})'\widehat{\mathbf{W}}(\mathbf{S}^{*}_{vy^{*}}-\mathbf{S}^{*}_{vz}\widehat{\boldsymbol{\theta}})$. The choice of $\mathbf{W}$ leads to a number of different GMM estimators, however, we will focus on the 2SLS weight matrix $(\mathbf{W_{2SLS}}=N(\mathbf{V}'\mathbf{V})^{-1})$. \citet{hayashi2011} provides a general treatment of the various single and multiple equation GMM estimators associated with $\mathbf{W}$ and \citet{bollen2014} develop a MIIV-GMM estimator for latent variable models.Under this framework we write the estimating equations for consistent estimation of the factor loadings and regression coefficients ($\boldsymbol{\theta}_{B\Lambda}$) from \eqref{full1} and \eqref{full2} as
\begin{IEEEeqnarray}{rCl} 
\hat{\boldsymbol{\theta}}_{B\Lambda} = 
(\hat{\mathbf{S}}^{*'}_{vz}\hat{\mathbf{S}}^{*-1}_{vv}\hat{\mathbf{S}}^{*}_{vz})^{-1}
 \hat{\mathbf{S}}^{*'}_{vz}\hat{\mathbf{S}}^{*-1}_{vv}\hat{\mathbf{S}}^{*}_{vy^{*}}.
\end{IEEEeqnarray}
\noindent where $\mathbf{S}^{*}_{vz}$ is the block-diagonal matrix containing the covariances between the explanatory variables and instruments, $\mathbf{S}^{*}_{vy^{*}}$ is the block-diagonal matrix containing the covariances between the explanatory variables and dependent variables, and $\mathbf{S}^{*}_{vv}$ is the block-diagonal instrument covariance matrix. \citet[Eq. 22]{fisher2019} provides formulas for consistent estimation of $\boldsymbol{\theta}_{B\Lambda}$  in the presence of equality constraints. With $\hat{\boldsymbol{\theta}}_{B\Lambda}$ in hand we can obtain consistent estimates of the latent variable and non-scaling indicator intercepts $\boldsymbol{\theta}_{\alpha}$, from \eqref{full1} and \eqref{full2} as $\hat{\boldsymbol{\theta}}_{\alpha}  =  \hat{\boldsymbol{\mu}}^{*'}_{y} -\hat{\boldsymbol{\mu}}^{*'}_{z}\hat{\boldsymbol{\theta}}_{B\Lambda}$. Combining these estimates gives us $\hat{\boldsymbol{\theta}}_{1}=[\hat{\boldsymbol{\theta}}_{B\Lambda},\hat{\boldsymbol{\theta}}_{\alpha}]$.
\subsubsection{Estimation of PIV Variance Parameters}

Estimation of the PIV variance parameters requires an expression for the model-implied variances of $\mathbf{y}^{*}$,
\begin{IEEEeqnarray}{rClCl} 
\boldsymbol{\Sigma}(\boldsymbol{\theta}) &=& 
\boldsymbol{\Lambda}(\mathbf{I}-\mathbf{B})^{-1}
\boldsymbol{\Sigma}_{\zeta}
(\mathbf{I}-\mathbf{B}^{'})^{-1}\boldsymbol{\Lambda}^{'}+
\boldsymbol{\Sigma}_{\varepsilon}
\end{IEEEeqnarray}
\noindent with $\boldsymbol{\Sigma}(\boldsymbol{\theta})$ representing the restrictions imposed on $\boldsymbol{\Sigma}^{*}$ by the parameters in $\boldsymbol{\theta}$. Similarly, we use $\boldsymbol{\gamma}(\boldsymbol{\theta}_{1})$ to denote the specific restrictions placed on $\upsilon({\boldsymbol{\Sigma}}^{*})$ by the PIV regression model parameters contained in $\boldsymbol{\theta}_{1}$. With consistent estimates of $\boldsymbol{\theta}_{1}$ in hand we can now 
estimate the parameters in $\boldsymbol{\theta}_{2}$ conditional on $\hat{\boldsymbol{\theta}}_{1}$. A number of system-wide estimators could be used to estimate $\boldsymbol{\theta}_{2}$. Here, we use a weighted least squares discrepancy function given by
\begin{IEEEeqnarray}{rClCl} 
\label{pivvar}
F(\boldsymbol{\theta};\:\mathbf{W}) &=& 
(\boldsymbol{\sigma}-\boldsymbol{\sigma}(\boldsymbol{\theta}_{2},\hat{\boldsymbol{\theta}}_{1}))^{'}
\hat{\mathbf{W}}_{WLS}
(\boldsymbol{\sigma}-\boldsymbol{\sigma}(\boldsymbol{\theta}_{2},\hat{\boldsymbol{\theta}}_{1}))
\end{IEEEeqnarray}
\noindent where 
$\boldsymbol{\sigma}(\boldsymbol{\theta}_{2},\hat{\boldsymbol{\theta}}_{1})$ represent the mean and variance parameters implied by the full model conditional on $\hat{\boldsymbol{\theta}}_{1}$ and $\hat{\mathbf{W}}_{WLS} = \mathbb{V}ar(\boldsymbol{\pi})^{-1}$ \citep{muthen1978}. These conditional variance and covariance parameter estimates will be consistent estimates of their population values, however, the naive standard errors corresponding to the estimates will be too small.  In fact, the standard errors for both $\boldsymbol{\theta}_{1}$ and $\boldsymbol{\theta}_{2}$  require modification and these adjustments are addressed now.\\ 
\subsubsection{Asymptotic Variance of $\boldsymbol{\theta}_{1}$}
A naive estimator for $\mathbb{V}ar(\boldsymbol{\theta}_{1})$ is $N^{-1}\mathbf{u^{'}u}
(\boldsymbol{\Sigma}^{*'}_{vz}\boldsymbol{\Sigma}^{*-1}_{vv}\boldsymbol{\Sigma}^{*}_{vz})^{-1} $ where $\mathbb{V}ar(\mathbf{u^{'}u})$ is estimated from the residual variances of the estimating equations. In the current context this estimator is not correct as it fails to account for the uncertainty arising in the estimation of  $\boldsymbol{\Sigma}^{*}$. In the context of endogenous ordinal variables \citet{bollen2007} show that $
\sqrt{N}(\hat{\boldsymbol{\theta}}_{1}-\boldsymbol{\theta}_{1})\xrightarrow[]{d} \mathcal{N}(\mathbf{0}, 
\mathbf{K}^{'}\:\mathbb{V}ar(\boldsymbol{\Sigma}^{*})\:\mathbf{K})$
where $\mathbf{K} = \partial\:\boldsymbol{\gamma}(\boldsymbol{\Sigma}^{*}) /
\partial\: \boldsymbol{\Sigma}^{*}$. Writing this quantity with respect to the patterning observed in the MIIV matrices we obtain the following approximation of the large sample covariance matrix of $\boldsymbol{\theta}_{1}$
\begin{IEEEeqnarray}{rClCl} 
\label{pivvartheta1}
\mathbb{V}ar(\hat{\boldsymbol{\theta}}_{1}) &=& \left.\frac{\partial\:\boldsymbol{\gamma}(\boldsymbol{\Sigma}^{*}) } 
{\partial\: \upsilon(\boldsymbol{\Sigma}^{*})}\right|^{'}_{ \boldsymbol{\Sigma}^{*}= \hat{\boldsymbol{\Sigma}}^{*}}
\mathbb{V}ar(\hat{\boldsymbol{\Sigma}}^{*})
 \left.\frac{\partial\:\boldsymbol{\gamma}(\boldsymbol{\Sigma}^{*}) } 
{\partial\: \upsilon(\boldsymbol{\Sigma}^{*})}\right|_{ \boldsymbol{\Sigma}^{*}= \hat{\boldsymbol{\Sigma}}^{*}}.
\end{IEEEeqnarray}
 \citet{bollen2007} only discussed ordinal variables and used numerical derivatives to obtain $\mathbf{K}$. Here we present the complementary analytic derivatives for each of the requisite submatrices required to obtain this quantity. Our derivatives are valid for a model where $\boldsymbol{\Sigma}^{*}_{vv}$ is strictly lower-triangular (ordinal variables only under the standard parameterization), a combination of strictly lower-triangular and lower-triangular (a combination of ordinal and continuous variables under the standard parameterization), or lower-triangular (any combination of variables under the alternative parameterization). The submatrices of $\mathbf{K}$ pertaining to the covariance elements are
\begin{IEEEeqnarray}{rClCl} 
\frac{\partial\:\boldsymbol{\gamma}(\boldsymbol{\theta}_{1})}
{\partial\: \mathrm{vec}(\boldsymbol{\Sigma}^{*}_{vz})^{'}}
& = & 
\big[(\boldsymbol{\Sigma}^{*'}_{vy}
\boldsymbol{\Sigma}^{*-1'}_{vv} \otimes \mathbf{U}^{-1})-
(\boldsymbol{\theta_{1}}^{'}\boldsymbol{\Sigma}^{*'}_{vz}
\boldsymbol{\Sigma}^{*-1'}_{vv} \otimes
\mathbf{U}^{-1})\big]\mathbf{K}_{s,r} -
(\boldsymbol{\theta_{1}}^{'} \otimes
\mathbf{U}^{-1} \boldsymbol{\Sigma}^{*'}_{vz}
\boldsymbol{\Sigma}^{*-1}_{vv})  \nonumber\\
\label{pd3_1e}
\frac{\partial\:\boldsymbol{\gamma}(\boldsymbol{\theta}_{1})}
{\partial\: \mathrm{vec}(\boldsymbol{\Sigma}^{*}_{vy})^{'}}
&=&  (
\boldsymbol{\Sigma}^{*'}_{vz}
\boldsymbol{\Sigma}^{*-1}_{vv}
\boldsymbol{\Sigma}^{*-1}_{vz})
\boldsymbol{\Sigma}^{*'}_{vz}
\boldsymbol{\Sigma}^{*-1}_{vv}
 \nonumber\\
\label{pd2_all}
\frac{\partial\:\boldsymbol{\gamma}(\boldsymbol{\theta}_{1})}{\partial\: 
\upsilon(\boldsymbol{\Sigma}^{*}_{vv})^{'}}
&=& 
\big[
(\mathbf{V}^{'}\mathbf{U}^{-1'}
\boldsymbol{\Sigma}^{*'}_{vz}
\boldsymbol{\Sigma}^{*-1'}_{vv}\otimes
\mathbf{U}^{-1}
\boldsymbol{\Sigma}^{*'}_{vz}
\boldsymbol{\Sigma}^{*-1}_{vv}) -
(\boldsymbol{\Sigma}^{*'}_{vy}
\boldsymbol{\Sigma}^{*-1'}_{vv} \otimes \mathbf{U}^{-1}
\boldsymbol{\Sigma}^{*'}_{vz}
\boldsymbol{\Sigma}^{*-1}_{vv})
\big]
\boldsymbol{\Delta} \nonumber
\end{IEEEeqnarray}
\noindent where $\mathbf{U} =  \boldsymbol{\Sigma}^{*'}_{vz}\boldsymbol{\Sigma}^{*-1}_{vv}\boldsymbol{\Sigma}^{*}_{vz}$ and $\mathbf{V} = \boldsymbol{\Sigma}^{*'}_{vz}\boldsymbol{\Sigma}^{*-1}_{vv}\boldsymbol{\Sigma}^{*}_{vy}$ and again, $\boldsymbol{\Sigma}^{*}_{vv}$ may take on an arbitrary patterning depending on the variable types and parameterization.\\

As  \citet{bollen2007} only considered ordinal variables they had no need for the mean structure of $\mathbf{y}^{*}$. Here we consider the estimation of mean structures for both continuous and ordinal variables and thus require results for these parameters as well. As the intercept parameters are dependent on $\mathbf{S}^{*}_{vz}$, $\mathbf{S}^{*}_{vy^{*}}$, $\mathbf{S}^{*}_{vv}$ and the means of the dependent, $\boldsymbol{\mu}^{*}_{y}$, and explanatory variables, $\boldsymbol{\mu}^{*}_{z}$, the partial derivatives with respect to each of these quantities is required. For the mean vectors the partial derivatives are straightforward, $\frac{\partial\:\boldsymbol{\gamma}(\boldsymbol{\mu}^{*})}
{\partial\: \boldsymbol{\mu}^{*'}_{y}}=\mathbf{I} $ and$ \frac{\partial\:\boldsymbol{\gamma}(\boldsymbol{\mu}^{*})}
{\partial\: \boldsymbol{\mu}^{*'}_{z}} = -\boldsymbol{\theta}_{B\Lambda}$.  The partial derivatives with respect to the matrices needed for the estimate of  $\boldsymbol{\theta}_{B\Lambda}$ requires building on previous results
\begin{IEEEeqnarray}{rClCl} 
\frac{\partial\:\boldsymbol{\gamma}(\boldsymbol{\theta}_{1})}
{\partial\: \mathrm{vec}(\boldsymbol{\Sigma}^{*}_{vz})^{'}}
& = & 
- \boldsymbol{\mu}^{*}_{z}
\frac{\partial\:\boldsymbol{\gamma}(\boldsymbol{\theta}_{1})}
{\partial\: \mathrm{vec}(\boldsymbol{\Sigma}^{*}_{vz})^{'}}  \nonumber\\
\frac{\partial\:\boldsymbol{\gamma}(\boldsymbol{\theta}_{1})}
{\partial\: \mathrm{vec}(\boldsymbol{\Sigma}^{*}_{vy})^{'}}
& = & 
- \boldsymbol{\mu}^{*}_{z}
\frac{\partial\:\boldsymbol{\gamma}(\boldsymbol{\theta}_{1})}
{\partial\: \mathrm{vec}(\boldsymbol{\Sigma}^{*}_{vy})^{'}}  \nonumber\\
\frac{\partial\:\boldsymbol{\gamma}(\boldsymbol{\theta}_{1})}
{\partial\: \upsilon(\boldsymbol{\Sigma}^{*}_{vv})^{'}}
& = & 
- \boldsymbol{\mu}^{*}_{z}
\frac{\partial\:\boldsymbol{\gamma}(\boldsymbol{\theta_{1}}^{*})}
{\partial\: \upsilon(\boldsymbol{\Sigma}^{*}_{vv})^{'}}  
\boldsymbol{\Delta},\nonumber
\end{IEEEeqnarray}
where again, the duplication matrix $\boldsymbol{\Delta}$ accounts for the arbitrary patterning of $\boldsymbol{\Sigma}^{*}_{vv}$.  Now, let $M_{1}=[\frac{\partial\:\boldsymbol{\gamma}(\boldsymbol{\theta}_{1})}
{\partial\: \mathrm{vec}(\boldsymbol{\Sigma}^{*}_{vz})^{'}},\frac{\partial\:\boldsymbol{\gamma}(\boldsymbol{\theta}_{1})}{\partial\: \mathrm{vec}(\boldsymbol{\Sigma}^{*}_{vy})^{'}},\frac{\partial\:\boldsymbol{\gamma}(\boldsymbol{\theta}_{1})}{\partial\: \upsilon(\boldsymbol{\Sigma}^{*}_{vv})^{'}}]$ and $M_{2}=$ $[\frac{\partial\:\boldsymbol{\gamma}(\boldsymbol{\theta}_{1})}{\partial\: \boldsymbol{\mu}^{*'}_{y}},$ $ \frac{\partial\:\boldsymbol{\gamma}(\boldsymbol{\theta}_{1})}{\partial\: \boldsymbol{\mu}^{*'}_{z}}, $ $\frac{\partial\:\boldsymbol{\gamma}(\boldsymbol{\theta}_{1})}
{\partial\: \mathrm{vec}(\boldsymbol{\Sigma}^{*}_{vz})^{'}}, $ $\frac{\partial\:\boldsymbol{\gamma}(\boldsymbol{\theta}_{1})}
{\partial\: \mathrm{vec}(\boldsymbol{\Sigma}^{*}_{vy})^{'}}, $ $\frac{\partial\:\boldsymbol{\gamma}(\boldsymbol{\theta}_{1})}
{\partial\: \upsilon(\boldsymbol{\Sigma}^{*}_{vv})^{'}}]$, and the partial derivatives of $\boldsymbol{\gamma}(\hat{\boldsymbol{\theta}}_{1})$ with respect to $\boldsymbol{\sigma}$ is
$\frac{\partial\:\boldsymbol{\gamma}(\hat{\boldsymbol{\theta}}_{1})} 
{\partial\: \boldsymbol{\sigma}} =
[\mathbf{M}_{1}^{'},\mathbf{M}_{2}^{'}]^{'}$ where the entries of $\mathbf{M}_1$ and $\mathbf{M}_2$ have been reordered to match the ordering of elements in  $\mathbb{V}ar(\hat{\boldsymbol{\sigma}}^{*})$.
\subsubsection{Asymptotic Variance of $\boldsymbol{\theta}_{2}$}

With a consistent approximation for the asymptotic distribution of $\boldsymbol{\theta}_{1}$ we employ results from
\citet[Eq. 31]{bollen2007} showing 
\begin{IEEEeqnarray}{rClCl} 
\label{pivvartheta2}
\mathbb{V}ar(\hat{\boldsymbol{\theta}}_{2}) &=&
\hat{\mathbf{H}}(\mathbf{I}-\hat{\mathbf{J}}_{1}\hat{\mathbf{K}})
\mathbb{V}ar(\hat{\boldsymbol{\Sigma}}^{*})
(\mathbf{I}-\hat{\mathbf{J}}_{1}\hat{\mathbf{K}})^{'}\hat{\mathbf{H}}^{'}
\end{IEEEeqnarray}
\noindent where $\mathbf{J}_{1}=\partial{\boldsymbol{\sigma}(\boldsymbol{\theta}_{1})} / \partial \boldsymbol{\theta}_{1}$, $\mathbf{H}=(\mathbf{J}_{2}\mathbf{J}_{2}^{'})^{-1}\mathbf{J}_{2}^{'}$, and $\mathbf{J}_{2}=\partial{\boldsymbol{\sigma}(\boldsymbol{\theta}_{2},\boldsymbol{\theta}_{1})} / \partial \boldsymbol{\theta}_{2}$, are all evaluated at $\hat{\boldsymbol{\theta}}$. For completeness the elements of $\mathbf{J}_{1}$ and $\mathbf{J}_{2}$ are also presented as matrix expressions to facilitate computational implementation of the estimator described,
\begin{IEEEeqnarray}{rClCl} 
\frac{\partial{\boldsymbol{\sigma}({\boldsymbol{\theta}}_{1})}}
{\partial\: \upsilon(\boldsymbol{\Lambda})^{'}}
& = & 
\boldsymbol{\Delta}^{+}
[(\boldsymbol{\Lambda}\mathbf{F}\boldsymbol{\Psi}\mathbf{F}^{'}\otimes\mathbf{I})
+ (\mathbf{I} \otimes \boldsymbol{\Lambda}\mathbf{F}\boldsymbol{\Psi}\mathbf{F}^{'})\mathbf{K}
]\boldsymbol{\Delta}, \\
\frac{\partial{\boldsymbol{\sigma}({\boldsymbol{\theta}}_{1})}}
{\partial\: \upsilon(\mathbf{B})^{'}}
& = & 
\boldsymbol{\Delta}^{+}
[(\boldsymbol{\Lambda}\mathbf{F}\boldsymbol{\Psi}\mathbf{F}^{'}\otimes\boldsymbol{\Lambda}\mathbf{F})
+ (\boldsymbol{\Lambda}\mathbf{F}\otimes \boldsymbol{\Lambda}\mathbf{F}\boldsymbol{\Psi}\mathbf{F}^{'})\mathbf{K}
]\boldsymbol{\Delta},\\
\frac{\partial{\boldsymbol{\sigma}(\boldsymbol{\theta}_{2},\hat{\boldsymbol{\theta}}_{1})}}
{\partial\: \upsilon(\boldsymbol{\Sigma}_{\zeta})^{'}}
& = & 
\boldsymbol{\Delta}^{+}(\boldsymbol{\Lambda}\mathbf{F} \otimes\boldsymbol{\Lambda}\mathbf{QF})
\boldsymbol{\Delta},
\\
\frac{\partial{\boldsymbol{\sigma}(\boldsymbol{\theta}_{2},\hat{\boldsymbol{\theta}}_{1})}}
{\partial\: \upsilon(\boldsymbol{\Sigma}_{\varepsilon})^{'}}
& = & 
\boldsymbol{\Delta}^{+}\boldsymbol{\Delta},
\end{IEEEeqnarray}
\noindent where $\mathbf{F} = (\mathbf{I}-\mathbf{B})^{-1}$. It should be noted the patterning of the elements in $\boldsymbol{\Lambda}$, $\mathbf{B}$, $\boldsymbol{\Sigma}_{\zeta}$ and $\boldsymbol{\Sigma}_{\varepsilon}$ is arbitrary and will not be known in advance. For this reason the algorithm described by \citet{nel1980} is required. Moreover, in the case of $\partial{\boldsymbol{\sigma}(\boldsymbol{\theta}_{2},\hat{\boldsymbol{\theta}}_{1})} / \partial\: \upsilon(\boldsymbol{\Sigma}_{\varepsilon})^{'}$, $\boldsymbol{\Delta}^{+}\boldsymbol{\Delta}\neq \mathbf{I}$, as $\boldsymbol{\Delta}^{+}$ is the elimination matrix associated with the model implied variances and covariances, and $\boldsymbol{\Delta}$ is the duplication matrix associated with the patterning specific to $\boldsymbol{\Sigma}_{\varepsilon}$. \\
 \begin{table}[ht]
        \centering
        \caption{\textcolor{black}{Description of estimation algorithm.}}
\scalebox{0.7}{
  \begin{threeparttable}
        \begin{tabular}{lcl}
        \toprule
        Parameterization & Step & Details\\
        \midrule
        Standard and Alternative &1a & Obtain consistent estimates of the sample means, thresholds, variances, covariances and \\ &&correlations, $\hat{\boldsymbol{\omega}} = [\hat{\boldsymbol{\mu}}_{y^{*}}, \hat{\boldsymbol{\tau}},\upsilon(\hat{\boldsymbol{\Sigma}}_{y^{*}})]$, where $\boldsymbol{\Sigma}_{y^{*}}$ is given in \eqref{sigystar} and the patterned matrix\\
        &&  operator $\upsilon(\cdot)$ is defined previously. \\
        &1b & Obtain an estimate of the large sample variances of $\boldsymbol{\omega}$ from Step 1a, $\mathbb{V}ar(\boldsymbol{\omega})$. \\
        \midrule
        Alternative Only & 2a & Obtain consistent estimates of the unconstrained mean vector, transformed thresholds,  \\ && unconstrained variances and covariances, $\hat{\boldsymbol{\pi}} = [\hat{\ddot{\boldsymbol{\tau}}}, \hat{\ddot{\boldsymbol{\mu}}}_{y^{*}},\upsilon(\hat{\ddot{\boldsymbol{\Sigma}}}_{y^{*}})]$, where the elements of \\
        && $\boldsymbol{\pi}$ are given in \eqref{taudot}-\eqref{sigdot}.\\
        &2b & Obtain an estimate of the large sample variances of $\boldsymbol{\pi}$ from Step 2a, $\mathbb{V}ar(\boldsymbol{\pi})$. \\
        \midrule
        Standard and Alternative & 3a & Obtain consistent estimates of the intercepts, factor loadings and regression parameters \\ && in $\boldsymbol{\theta}_{1}$, $\hat{\boldsymbol{\theta}}_{1} = [\upsilon(\hat{\boldsymbol{\alpha}}_{\eta}), \upsilon( \hat{\boldsymbol{\alpha}}_{y^{*}}),\upsilon( \boldsymbol{\Lambda}),\upsilon( \mathbf{B})]$, as described in \eqref{pivtheta1a}.\\
                & 3b & Obtain consistent estimates of the variance parameters in $\boldsymbol{\theta}_{2}$, where $\hat{\boldsymbol{\theta}}_{2} = [\hat{\boldsymbol{\Sigma}}_{\zeta}, \hat{\boldsymbol{\Sigma}}_{\varepsilon}]$, as \\ && given in \eqref{pivvar}.\\
                & 3c & Obtain estimates of the large sample variance of the estimates obtained in 3a, $\mathbb{V}ar(\boldsymbol{\theta}_{1})$ \\ && given in \eqref{pivvartheta1}.\\
               & 3d & Obtain estimates of the large sample variance of the estimates obtained in 3b, $\mathbb{V}ar(\boldsymbol{\theta}_{2})$ \\ && given in \eqref{pivvartheta2}.\\
               \bottomrule
        \end{tabular}
           \begin{tablenotes}
 \end{tablenotes}
   \end{threeparttable}
}
\label{algo}
\end{table}
\section{An Illustrative Example}

\begin{figure}[h!]
  \includegraphics[width=.8\textwidth]{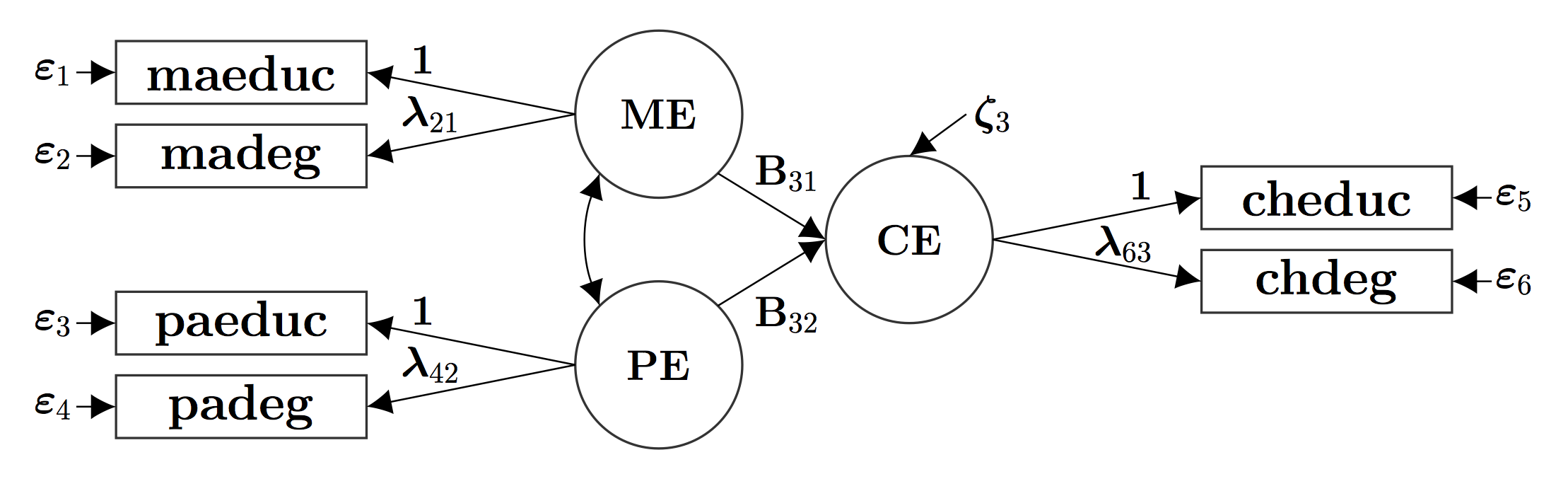}
  \caption{Path Diagram for Empirical Example}
  \label{fig1}
\end{figure}

\begin{table}[ht]
\centering
\caption{Empirical Example} 
\scalebox{0.7}{
  \begin{threeparttable}
\begin{tabular}{rrrrrr}
  \toprule
  &    \multicolumn{5}{c}{Estimator}  \\
 \cmidrule(lr){2-3}
   \cmidrule(lr){4-6} &    \multicolumn{2}{c}{WLSMV} &
        \multicolumn{3}{c}{MIIV} 
  \\
 \cmidrule(lr){2-3}
   \cmidrule(lr){4-6} \multicolumn{1}{c}{Parameter} &
  \multicolumn{1}{c}{Est.} &
  \multicolumn{1}{c}{Std.Err.} &
  \multicolumn{1}{c}{Est.} &  
    \multicolumn{1}{c}{Std.Err.} &
  \multicolumn{1}{c}{$R^{2}_{S}$}\\
 \midrule
$\lambda_{(1,1)}$ & 1.00 & & 1.00 & &  \\ 
  $\lambda_{(2,1)}$ & 0.70 & 0.02 & 0.67 & 0.02 & 0.49 \\ 
    $\lambda_{(3,2)}$ & 1.00 &  & 1.00 &  &  \\ 
  $\lambda_{(4,2)}$ & 0.74 & 0.02 & 0.72 & 0.02 & 0.51 \\ 
  $\lambda_{(5,3)}$ & 1.00 &  & 1.00 &  &  \\ 
  $\lambda_{(6,3)}$ & 0.76 & 0.02 & 0.75 & 0.02 & 0.21 \\ 
  $\beta_{(3,1)}$ & 0.16 & 0.02 & 0.20 & 0.03 & 0.81 \\ 
  $\beta_{(3,2)}$ & 0.23 & 0.02 & 0.19 & 0.02 & 0.83 \\ 
  $\tau_{(2,1)}$ & 12.00 & 0.00 & 12.00 & 0.00 &  \\ 
  $\tau_{(2,2)}$ & 15.57 & 0.04 & 15.57 & 0.04 &  \\ 
  $\tau_{(2,3)}$ & 16.00 & 0.00 & 16.00 & 0.00 &  \\ 
  $\tau_{(2,4)}$ & 17.62 & 0.11 & 17.62 & 0.11 &  \\ 
  $\tau_{(4,1)}$ & 12.00 & 0.00 & 12.00 & 0.00 &  \\ 
  $\tau_{(4,2)}$ & 15.69 & 0.04 & 15.69 & 0.04 &  \\ 
  $\tau_{(4,3)}$ & 16.00 & 0.00 & 16.00 & 0.00 &  \\ 
  $\tau_{(4,4)}$ & 17.77 & 0.12 & 17.77 & 0.12 &  \\ 
  $\tau_{(6,1)}$ & 12.00 & 0.00 & 12.00 & 0.00 &  \\ 
  $\tau_{(6,2)}$ & 15.57 & 0.03 & 15.57 & 0.03 &  \\ 
  $\tau_{(6,3)}$ & 16.00 & 0.00 & 16.00 & 0.00 &  \\ 
  $\tau_{(6,4)}$ & 17.60 & 0.09 & 17.60 & 0.09 &  \\ 
  $\alpha_{\eta_{3}}$ & 9.59 & 0.20 & 9.51 & 0.21 &  \\ 
  $\alpha_{y_{2}}$ & 5.41 & 0.21 & 5.74 & 0.20 &  \\ 
  $\alpha_{y_{4}}$ & 4.94 & 0.24 & 5.18 & 0.24 &  \\ 
  $\alpha_{y_{6}}$ & 4.46 & 0.31 & 4.54 & 0.31 &  \\ 
  $\Sigma_{\zeta_{(1,1)}}$ & 12.56 & 0.41 & 12.82 & 0.41 &  \\ 
  $\Sigma_{\zeta_{(2,2)}}$ & 15.31 & 0.51 & 15.65 & 0.54 &  \\ 
  $\Sigma_{\zeta_{(3,3)}}$ & 6.19 & 0.24 & 6.34 & 0.26 &  \\ 
  $\Sigma_{\zeta_{(1,2)}}$ & 10.03 & 0.37 & 10.33 & 0.37 &  \\ 
  $\Sigma_{\varepsilon_{(1,1)}}$ & 1.10 & 0.13 & 0.83 & 0.15 &  \\ 
  $\Sigma_{\varepsilon_{(2,2)}}$ & 0.21 & 0.07 & 0.42 & 0.08 &  \\ 
  $\Sigma_{\varepsilon_{(3,3)}}$ & 1.07 & 0.15 & 0.73 & 0.23 &  \\ 
  $\Sigma_{\varepsilon_{(4,4)}}$ & 0.21 & 0.09 & 0.31 & 0.13 &  \\ 
  $\Sigma_{\varepsilon_{(5,5)}}$ & 0.60 & 0.17 & 0.40 & 0.18 &  \\ 
  $\Sigma_{\varepsilon_{(6,6)}}$ & 0.38 & 0.10 & 0.52 & 0.11 &  \\ 
    \bottomrule
  \end{tabular}
      \begin{tablenotes}
    \item Note. For the \emph{padeg} equation, all items except for \emph{paeduc} were used as MIIVs. For \emph{madeg}, all  items except for \emph{maeduc} were used as MIIVs. For \emph{chdeg}, all items except for \emph{cheduc} were used as MIIVs. For the \emph{CE} equation, all items except for  \emph{cheduc} and  \emph{chdeg} served as MIIVs.  $R^2_S$ indicates Shea's partial $R^2$, an index of instrument strength.
 \end{tablenotes}
   \end{threeparttable}
}
\label{gsstable}
\end{table}

\textcolor{black}{To demonstrate the PIV estimator developed here and the utility of our general alternative parameterization we used data from the 2016 General Social Survey (GSS). We chose to model three constructs representing educational attainment; maternal education (ME), paternal education (PE) and the respondent (or child's) education (CE). Each construct was measured using two indicators, a near continuous variable measuring years of education (educ), and an ordered categorical variable representing the highest degree achieved (deg).  Response options for the degree question were: (1) \emph{less than high school}, (2) \emph{high school}, (3) \emph{junior college}, (4) \emph{bachelor}, and (5) \emph{graduate}. We hypothesized that \emph{high school} and \emph{bachelor} degrees were the response options linked most closely to consistent durations, at $12$ and $16$ years, respectively.  For this reason we fixed these two thresholds to the numerical values $12$ and $16$ and freely estimated the thresholds for \emph{junior college} and \emph{graduate} degrees.  Furthermore, we regressed child's educational attainment on the parent's educational attainment. See Figure~\ref{fig1} for a path diagram of the fitted model.} \\
\textcolor{black}{To obtain the PIV estimates for the model shown in Figure~\ref{fig1} we used the \texttt{MIIVsem} package \citep{fisher2017}. For comparative purposes we also fit the model using the WSLMV estimator in \texttt{lavaan} \citep{rosseel2012}. The data contained observations from $1,910$ subjects. Estimates from the two estimators are provided in Table~\ref{gsstable}.} 
\textcolor{black}{An examination of the threshold estimates paints an interesting picture in terms of the correspondence between the years of education associated with obtaining a junior college and graduate degree.  For example, in this sample a junior college degree is associated with approximately $15.6$ years of schooling while a graduate degree is associated with approximately $17.7$  years. Importantly, a high degree of consistency emerged in these estimates across the mother, father and child's education.  Overall, the factor loadings obtained from the PIV estimator are trivially smaller than those from WLSMV estimator. A similar pattern does not emerge for the regression coefficients, intercepts and variance parameters. Standard errors for the two estimators are comparable across the parameter sets and inferences based on the two estimators are also in agreement.  Furthermore, Shea's partial $R^2$ measures for the structural parameters in our model are sufficiently large to avoid concerns of weak instruments. Additional indirect evidence of instrument strength is provided by the consistency across the two estimators and this is typically also an indication of a well-specified model.}

\section{Simulation Study}

In this section we examine the finite sample properties of the proposed point estimates and their standard errors in a small simulation study.  To promote reproducibility and enrich cross-study comparisons we adapt the main simulation condition from \citet{jin2016} to the case of mixed ordinal and continuous endogenous variables. MIIV estimates were obtained using the \texttt{MIIVsem} package \citep{fisher2017} and  WSLMV estimates using \texttt{lavaan} \citep{rosseel2012}. 

\subsection{Model Specification and Data Generation}

\textcolor{black}{The following data generating parameters were employed across all simulation conditions: $[B_{(3,1)}, B_{(4,2)},B_{(5,2)},B_{(5,3)},B_{(5,4)}]=[0.5,0.4,0.3,0.4,0.4]$, $[\lambda_{(1,1)},$ $\lambda_{(2,1)},$ $\lambda_{(3,2)},$ $\lambda_{(4,2)}, $ $\lambda_{(5,2)}, $ $\lambda_{(6,3)}, $ $ \lambda_{(7,3)}, $ $\lambda_{8,3},$ $\lambda_{(9,4)},$ $\lambda_{(10,4)},$ $\lambda_{(11,5)},$ $\lambda_{(12,5)}]$ $ =$ $[1.0,$ $0.4,$ $1.0,$ $0.7, $ $0.6,$ $1.0,$ $0.8,$ $0.7,$ $1.0,$ $0.6,$ $1.0,$ $0.5]$, and $[\Sigma_{\zeta_{(1,1)}},$ $\Sigma_{\zeta_{(2,1)}},$ $\Sigma_{\zeta_{(2,2)}},$ $\Sigma_{\zeta_{(3,3)}},$ $\Sigma_{\zeta_{(4,4)}},$ $\Sigma_{\zeta_{(5,5)}}$ $]=[0.7,$ $0.3,$ $0.8,$ $0.4,$ $0.5,$ $0.5]$. As we discuss in the results, these parameter values led to a range of strengths of the instruments so we can look at the impact of weak instrumental variables.  Of the 5 latent variables included in the analysis, $\eta_{1}$ and $\eta_{2}$ were measured by the continuous indicators $y_{1}-y_{2}$ and $y_{3}-y_{5}$, respectively, while  $\eta_{3}$, $\eta_{4}$ and $\eta_{5}$ were measured by the 5-category ordinal variables $y_{6}-y_{8}$, $y_{9}-y_{10}$ and $y_{11}-y_{12}$, respectively. All observed variables were generated according to a normal distribution with mean zero and subsequently $y_{6}-y_{12}$ were discretized with response probabilities of $0.3,$ $0.4,$ $0.2,$ $0.06,$ $0.04$. Data were generated according to five sample sizes ($N=100,200,400,800,3200$) commonly encountered in empirical research and for each sample size $2,000$ datasets were generated.}  
 
\subsection{Improper Solutions and Nonconvergence}

For the purposes of our analysis nonpositive definite factor and error variance matrices were flagged. Previous examinations of the PIV estimator have handled improper solutions differently.  For example, \citet{nestler2013} retained Heywood cases for all estimators in the final analysis.  The rationale here is that Heywood cases reflect true sampling variability of the estimator and through their omission bias is introduced in summary statistics involving means or variances.   On the other hand \citet{jin2016} omitted solutions containing nonpositive definite matrices from all summary statistics except for the calculation of a parameter's empirical standard deviation, which was subsequently used to asses the accuracy of estimated standard errors.  For this reason we have chosen to conduct a sensitivity analysis to better understand the impact of omitting these solutions from our analysis.  Across all simulation conditions solutions which did not converge were omitted from the analysis.  This is discussed further in the results section and a breakdown of the percentage of nonpositive definite matrices and nonconverged solutions by estimator and parameterization is given in Table~\ref{baddata}.
\vspace{-.4cm}
\subsection{Outcome Variables}
\vspace{-.4cm}
In line with previous simulations we examined the relative bias of both the point estimates and standard errors within each simulation condition. The mean percentage of relative bias for point estimates was calculated as $RB=\mathrm{mean}[(\hat{\theta}_{ak}-\theta_{a})/\theta_{a}]100$ where $\theta_{a}$ is the data generating parameter in a given simulation condition and $\hat{\theta}_{ak}$ is the estimate for parameter $a$ in the $k^{th}$ Monte Carlo replication. We calculated the median percentage of relative bias for the standard errors as $RBSE$ $=$ $\mathrm{median}$ $[(SE(\hat{\theta}_{ak})$ $-$ $SD(\theta_{a}))$ $/SD(\theta_{a})$ $]100$. Here we use the heuristic that a relative bias of less than $|0.05|$ is considered trivial. The median was chosen to ensure a robust measure of central tendency given the inclusion of improper solutions in our analysis.  Consistent with previous simulation studies, the mean RB and mean RBSE were calculated across each parameter set ($\boldsymbol{\tau}$, $\boldsymbol{\alpha}_{\eta}$, $\boldsymbol{\alpha}_{y^{(c)}}$, $\boldsymbol{\alpha}_{y^{(o)}}$, $\boldsymbol{\Lambda}_{y^{(c)}}$,  $\boldsymbol{\Lambda}_{y^{(o)}}$,$\mathbf{B}$, $\boldsymbol{\Sigma}_{\varepsilon_{y^{(c)}}}$,$\boldsymbol{\Sigma}_{\varepsilon_{y^{(o)}}}$, $\boldsymbol{\Sigma}_{\zeta}$) within a given simulation condition. Here a superscript of ${y^{(c)}}$ indicates parameters corresponding to one of the continuous variables in the system and a superscript of ${y^{(o)}}$ indicates parameters corresponding to an ordered categorical variable. \textcolor{black}{Finally, Shea's partial $R^2$ was used to assess instrument strength.}
\subsection{Results}
\begin{table}[ht]
\centering
\caption{Percentage of Nonconverging and Nonpositive Definite Solutions} 
\scalebox{0.7}{
\begin{tabular}{rcccc cccc}
  \toprule
  &    \multicolumn{8}{c}{Nonconverged or Nonpositive Definite (\%)}  \\
   \cmidrule(lr){2-9}
   & \multicolumn{4}{c}{Standard Parameterization} &
   \multicolumn{4}{c}{Alternative Parameterization} \\
   \cmidrule(lr){2-5} \cmidrule(lr){6-9} 
   &    \multicolumn{2}{c}{Noncoverged} &
        \multicolumn{2}{c}{Nonpositive Definite}  &
        \multicolumn{2}{c}{Noncoverged} &
        \multicolumn{2}{c}{Nonpositive Definite}  \\
 \cmidrule(lr){2-3}
   \cmidrule(lr){4-5} 
    \cmidrule(lr){6-7}
   \cmidrule(lr){8-9} 
   \multicolumn{1}{c}{Sample Size} &
     \multicolumn{1}{c}{MIIV} &
        \multicolumn{1}{c}{WLSMV} &
        \multicolumn{1}{c}{MIIV} &
        \multicolumn{1}{c}{WLSMV} &
        \multicolumn{1}{c}{MIIV} &
        \multicolumn{1}{c}{WLSMV} &
        \multicolumn{1}{c}{MIIV} &
        \multicolumn{1}{c}{WLSMV} \\
 \midrule
100 & 0.0 & 4.5 & 45.6 & 54.5 & 0.0 & 15.3 & 46.2 & 48.1 \\  
  200 & 0.0 & 0.3 & 24.1 & 26.6 & 0.0 & 0.8 & 23.7 & 26.6 \\ 
  400 & 0.0 & 0.0 & 9.3 & 7.4 & 0.0 & 0.0 & 9.0 & 7.4 \\ 
  800 & 0.0 & 0.0 & 2.9 & 1.7 & 0.0 & 0.0 & 2.8 & 1.7 \\ 
  3200 & 0.0 & 0.0 & 0.0 & 0.0 & 0.0 & 0.0 & 0.0 & 0.0 \\ 
  \bottomrule
  \end{tabular}
}
\label{baddata}
\end{table}

\begin{table}
 \label{sim1res}
\scalebox{0.62}{
  \begin{threeparttable}
    \caption{Simulation Results.}
\begin{tabular}{lrrrrrrrrrrrrrrrrrrrr}
  \toprule
  &    \multicolumn{20}{c}{Percentage of Relative Bias}  \\
 \cmidrule(lr){2-21} &    \multicolumn{10}{c}{Point Estimates} &
        \multicolumn{10}{c}{Standard Errors} 
  \\
 \cmidrule(lr){2-11}
   \cmidrule(lr){12-21} &    \multicolumn{10}{c}{Estimator}  &
        \multicolumn{10}{c}{Estimator}  
  \\
 \cmidrule(lr){2-11}
   \cmidrule(lr){12-21} &    \multicolumn{5}{c}{MIIV} &
        \multicolumn{5}{c}{WLSMV} &
        \multicolumn{5}{c}{MIIV} &
        \multicolumn{5}{c}{WLSMV} \\
 \cmidrule(lr){2-6}
  \cmidrule(lr){7-11}
  \cmidrule(lr){12-16}
  \cmidrule(lr){17-21} &    \multicolumn{5}{c}{Sample Size} &
        \multicolumn{5}{c}{Sample Size} &
        \multicolumn{5}{c}{Sample Size} &
        \multicolumn{5}{c}{Sample Size} \\
 \cmidrule(lr){2-6}
  \cmidrule(lr){7-11}
  \cmidrule(lr){12-16}
  \cmidrule(lr){17-21} \multicolumn{1}{c}{Parameter} &
  \multicolumn{1}{c}{100} &
  \multicolumn{1}{c}{200} &
  \multicolumn{1}{c}{400} &  
  \multicolumn{1}{c}{800} &
  \multicolumn{1}{c}{3,200} &
  \multicolumn{1}{c}{100} &
  \multicolumn{1}{c}{200} &
  \multicolumn{1}{c}{400} &  
  \multicolumn{1}{c}{800} &
  \multicolumn{1}{c}{3,200} &
  \multicolumn{1}{c}{100} &
  \multicolumn{1}{c}{200} &
  \multicolumn{1}{c}{400} &  
  \multicolumn{1}{c}{800} &
  \multicolumn{1}{c}{3,200} &
  \multicolumn{1}{c}{100} &
  \multicolumn{1}{c}{200} &
  \multicolumn{1}{c}{400} &  
  \multicolumn{1}{c}{800} &
  \multicolumn{1}{c}{3,200}\\
 \midrule
    \multicolumn{21}{c}{Standard Parameterization - Datasets with NPD Matrices Excluded} \\
  \midrule
  $\boldsymbol{\alpha}_{\eta}$ &  &  &  &  &  &  &  &  &  &  &  2.6 & -0.7 &  0.5 &  1.5 &  0.2 &   2.7 & -0.1 &  0.6 &  1.1 & -0.9 \\ 
$\boldsymbol{\tau}$ &  1.2 &  0.7 &  0.4 &  0.3 &  0.1 &   1.2 &  0.7 &  0.4 &  0.3 &  0.1 & \textbf{-5.8} & -2.5 & -1.5 & -0.1 &  0.4 & \textbf{ -5.8} & -2.5 & -1.5 & -0.1 &  0.4 \\ 
  $\boldsymbol{\Lambda_{y^(c)}}$ &  3.7 &  2.1 &  0.5 & -0.2 & -0.1 & \textbf{ 11.9} & \textbf{ 6.0} &  2.1 &  0.3 &  0.0 &  1.9 &  1.0 &  2.1 & -1.4 & -1.2 & \textbf{ -5.3} & -0.9 &  0.9 & -1.8 & -0.7 \\ 
  $\boldsymbol{\Lambda_{y^(o)}}$ & \textbf{-8.7} & -3.3 & -1.5 & -0.6 & -0.2 & \textbf{  8.9} &  4.5 &  1.3 &  0.7 &  0.1 &  0.0 &  0.1 & -0.9 & -0.5 & -0.5 & \textbf{-14.3} & \textbf{-6.3} & -2.7 & -2.3 & -0.4 \\
    $\mathbf{B}$ & -0.6 &  0.3 &  0.0 & -0.5 & -0.1 & \textbf{  9.6} & \textbf{ 5.4} &  2.4 &  0.7 &  0.2 &  1.8 &  1.4 &  0.6 &  0.6 &  0.6 & \textbf{-12.6} & \textbf{-9.1} & -4.0 & -2.1 & -0.1 \\ 
  $\boldsymbol{\Sigma}_{\varepsilon_{y^(c)}}$ & -0.2 & -1.0 & -1.0 & -1.5 & -0.5 & \textbf{ 10.8} &  5.0 &  1.9 & -0.2 & -0.1 & \textbf{13.7} & \textbf{ 6.6} &  4.0 &  0.7 & -2.6 &  -0.3 & -1.1 &  1.0 & -0.5 & -2.2 \\ 
  $\boldsymbol{\Sigma}_{\zeta}$ & \textbf{13.3} &  4.4 &  2.5 &  2.3 &  0.6 & \textbf{-10.6} & \textbf{-5.4} & -1.6 &  0.1 &  0.0 & \textbf{-5.3} &  3.3 &  2.0 &  1.0 & -0.1 & \textbf{ 11.2} &  2.7 & -0.8 & -2.2 & -1.2 \\ 
  \midrule
   \multicolumn{21}{c}{Standard Parameterization - All Converged Datasets} \\
  \midrule
  $\boldsymbol{\alpha}_{\eta}$ &  &  &  &  &  &  &  &  &  &  &   2.0 &  -0.8 &  0.2 &  1.1 &  0.2 &   0.9 &  -0.7 &  0.0 &  1.0 & -0.9 \\ 
    $\boldsymbol{\tau}$ &   1.4 &  0.9 &  0.4 &  0.3 &  0.1 &  1.4 &  0.9 &  0.4 &  0.3 &  0.1 &  -5.0 &  -3.1 & -1.5 & -0.2 &  0.4 &  -5.0 &  -3.1 & -1.5 & -0.2 &  0.4 \\ 
  $\boldsymbol{\Lambda_{y^{(c)}}}$ &  -3.0 & -1.8 & -0.7 & -0.5 & -0.1 &  4.7 &  2.2 &  1.0 &  0.0 &  0.0 &  -2.4 &  -2.7 &  0.4 & -1.9 & -1.2 & \textbf{-10.7} & \textbf{ -5.7} & -1.3 & -2.3 & -0.7 \\ 
  $\boldsymbol{\Lambda_{y^{(o)}}}$ & \textbf{-11.7} & -4.9 & -1.8 & -0.7 & -0.2 & \textbf{ 5.8} &  3.0 &  1.2 &  0.7 &  0.1 & \textbf{ -5.3} &  -4.0 & -2.1 & -0.8 & -0.5 & \textbf{-23.3} & \textbf{-11.2} & -4.3 & -2.9 & -0.4 \\ 
    $\mathbf{B}$ &  -3.5 & -1.8 & -0.8 & -0.8 & -0.1 & \textbf{ 8.7} &  3.7 &  1.8 &  0.5 &  0.2 & \textbf{-15.3} & \textbf{ -9.9} & -2.5 & -0.7 &  0.6 & \textbf{-35.7} & \textbf{-19.4} & \textbf{-5.8} & -2.9 & -0.1 \\ 
  $\boldsymbol{\Sigma}_{\varepsilon_{y^{(c)}}}$ & \textbf{-12.4} & \textbf{-9.4} & -4.2 & -2.4 & -0.5 & \textbf{-6.2} & -3.1 & -0.2 & -0.7 & -0.1 & \textbf{-12.6} & \textbf{ -9.0} & -3.7 & -2.3 & -2.6 & \textbf{-22.9} & \textbf{-20.8} & -3.6 & -2.2 & -2.2 \\ 
  $\boldsymbol{\Sigma}_{\zeta}$ & \textbf{ 26.0} & \textbf{ 9.7} &  4.9 &  2.9 &  0.6 & \textbf{19.3} &  2.1 &  0.2 &  0.5 &  0.0 & \textbf{-33.3} & \textbf{-24.6} & \textbf{-6.4} & -1.9 & -0.1 & \textbf{-51.6} & \textbf{-31.9} & \textbf{-8.1} & -4.6 & -1.2 \\ 
    \midrule
   \multicolumn{21}{c}{Alternative Parameterization - Datasets with NPD Matrices Excluded} \\
  \midrule
      $\boldsymbol{\alpha}_{\eta}$ & \textbf{16.0} & \textbf{ 7.8} &  2.3 &  2.0 &  0.6 & \textbf{ -6.9} & \textbf{-6.8} & -3.3 & -1.4 & -0.3 &   1.9 & -0.6 & -0.3 & -1.8 &  1.8 &  -3.8 & -4.4 & -1.6 & -2.8 &  1.5 \\ 
  $\boldsymbol{\alpha}_{y^{(o)}}$ & \textbf{20.3} & \textbf{ 7.1} &  2.3 &  0.8 &  0.3 & \textbf{-18.7} & \textbf{-9.7} & -3.7 & -2.1 & -0.1 &  -4.8 & -2.1 & -1.7 & -0.6 & -0.5 & \textbf{-17.5} & \textbf{-7.1} & -3.6 & -2.5 & -0.4 \\ 
$\boldsymbol{\tau}$ &  1.2 &  0.7 &  0.3 &  0.2 &  0.0 &   1.2 &  0.7 &  0.3 &  0.2 &  0.0 & \textbf{ -5.5} & -3.6 & -1.8 &  0.1 &  0.8 & \textbf{ -5.5} & -3.6 & -1.8 &  0.1 &  0.8 \\ 
  $\boldsymbol{\Lambda}_{y^{(c)}}$ &  4.5 &  2.0 &  0.5 & -0.2 & -0.1 & \textbf{ 12.6} & \textbf{ 6.0} &  2.0 &  0.3 &  0.0 &   2.2 &  1.4 &  2.1 & -1.3 & -1.2 &  -5.0 & -0.6 &  0.9 & -1.7 & -0.7 \\ 
  $\boldsymbol{\Lambda}_{y^{(o)}}$ & \textbf{-5.4} & -2.1 & -0.8 & -0.2 & -0.1 & \textbf{ 13.1} & \textbf{ 5.8} &  2.1 &  1.2 &  0.2 & \textbf{ -5.4} & -2.9 & -2.5 & -1.2 & -0.1 & \textbf{-17.3} & \textbf{-7.3} & -4.8 & -3.2 & -0.3 \\ 
    $\mathbf{B}$ &  1.7 &  1.2 &  0.4 & -0.4 & -0.1 & \textbf{ 12.4} & \textbf{ 6.4} &  2.8 &  0.9 &  0.2 &  -0.2 &  0.5 &  0.1 &  0.2 &  0.7 & \textbf{-15.4} & \textbf{-8.9} & -4.7 & -2.2 &  0.2 \\ 
  $\boldsymbol{\Sigma}_{\varepsilon_{y^{(c)}}}$ & -0.3 & -1.2 & -1.1 & -1.5 & -0.4 & \textbf{ 11.4} &  5.0 &  1.9 & -0.2 & -0.1 & \textbf{ 13.5} & \textbf{ 5.8} &  3.3 &  0.0 & -3.2 &  -0.7 & -1.1 &  1.1 & -0.5 & -2.2 \\ 
  $\boldsymbol{\Sigma}_{\varepsilon_{y^{(o)}}}$ & \textbf{-7.7} & -3.0 & -2.0 & -1.2 & -0.2 &   1.1 &  0.7 & -0.6 & -0.4 &  0.0 & \textbf{-13.9} & -4.6 & -1.6 & -0.4 &  0.1 & \textbf{ -6.0} & -4.1 & -1.5 & -1.8 & -0.6 \\ 
    $\boldsymbol{\Sigma}_{\zeta}$ & \textbf{14.1} & \textbf{ 5.8} &  2.8 &  2.3 &  0.6 & \textbf{ -8.4} & -4.0 & -1.1 &  0.3 &  0.1 &  -1.2 &  0.8 &  0.7 &  0.4 &  0.9 &   4.7 & -0.7 & -1.8 & -2.5 & -0.8 \\ 
      \midrule
   \multicolumn{21}{c}{Alternative Parameterization - All Converged Datasets} \\
  \midrule
    $\boldsymbol{\alpha}_{\eta}$ & \textbf{ 14.6} & \textbf{ 5.7} &  2.9 &  2.4 &  0.6 & \textbf{-13.0} & \textbf{-9.7} & -3.2 & -1.2 & -0.3 & \textbf{-11.3} & \textbf{ -9.5} & -0.9 & -1.8 &  1.8 & \textbf{-20.2} & \textbf{ -8.4} & -1.9 & -2.7 &  1.5 \\ 
  $\boldsymbol{\alpha}_{y^{(o)}}$ & \textbf{ 22.6} & \textbf{ 9.6} &  2.6 &  0.8 &  0.3 & \textbf{-18.4} & \textbf{-8.2} & -3.9 & -2.2 & -0.1 & \textbf{ -7.7} &  -4.3 & -2.4 & -0.8 & -0.5 & \textbf{-22.3} & \textbf{-10.4} & -4.7 & -2.8 & -0.4 \\ 
    $\boldsymbol{\tau}$ &   1.2 &  0.7 &  0.3 &  0.2 &  0.0 &   1.2 &  0.7 &  0.3 &  0.2 &  0.0 & \textbf{ -7.0} &  -3.6 & -1.6 &  0.1 &  0.8 & \textbf{ -7.0} &  -3.6 & -1.6 &  0.1 &  0.8 \\ 
  $\boldsymbol{\Lambda}_{y^{(c)}}$ &  -3.0 & -1.9 & -0.7 & -0.5 & -0.1 &   5.0 &  2.2 &  1.0 &  0.0 &  0.0 &  -3.0 &  -2.7 &  0.4 & -1.9 & -1.2 & \textbf{-11.3} & \textbf{ -5.7} & -1.3 & -2.3 & -0.7 \\ 
  $\boldsymbol{\Lambda}_{y^{(o)}}$ & \textbf{ -8.4} & -3.7 & -1.0 & -0.2 & -0.1 & \textbf{  9.8} &  4.3 &  2.0 &  1.2 &  0.2 & \textbf{ -8.5} & \textbf{ -5.1} & -2.9 & -1.5 & -0.1 & \textbf{-21.4} & \textbf{-10.0} & \textbf{-5.2} & -3.6 & -0.3 \\
    $\mathbf{B}$ &  -2.4 & -1.0 & -0.5 & -0.6 & -0.1 & \textbf{  9.8} &  4.7 &  2.1 &  0.7 &  0.2 & \textbf{-17.9} & \textbf{-10.6} & -2.3 & -0.9 &  0.7 & \textbf{-35.8} & \textbf{-15.5} & \textbf{-6.0} & -2.9 &  0.2 \\  
  $\boldsymbol{\Sigma}_{\varepsilon_{y^{(c)}}}$ & \textbf{-13.3} & \textbf{-9.6} & -4.2 & -2.4 & -0.4 &  -4.8 & -3.1 & -0.2 & -0.7 & -0.1 & \textbf{-13.8} & \textbf{ -9.8} & -4.4 & -3.0 & -3.2 & \textbf{-23.2} & \textbf{-20.9} & -3.6 & -2.2 & -2.2 \\ 
  $\boldsymbol{\Sigma}_{\varepsilon_{y^{(o)}}}$ & \textbf{-13.7} & -4.0 & -2.5 & -1.3 & -0.2 & \textbf{-10.1} & -2.0 & -1.0 & -0.5 &  0.0 & \textbf{-25.7} & \textbf{-18.5} & -3.7 & -0.9 &  0.1 & \textbf{-35.4} & \textbf{-16.6} & -4.6 & -2.7 & -0.6 \\ 
$\boldsymbol{\Sigma}_{\zeta}$ & \textbf{ 27.9} & \textbf{10.9} & \textbf{ 5.1} &  2.9 &  0.6 & \textbf{ 13.7} &  3.2 &  0.7 &  0.7 &  0.1 & \textbf{-33.7} & \textbf{-25.4} & \textbf{-7.5} & -2.2 &  0.9 & \textbf{-48.4} & \textbf{-31.6} & \textbf{-8.3} & -4.7 & -0.8 \\ 
  \bottomrule
  \end{tabular}
     \begin{tablenotes}
      \small
      \item \emph{Note.} $y^(c)$ represent parameters corresponding to continuous variables and $y^(o)$ represent parameters corresponding to ordered categorical variables. Relative bias could not be calculated for the latent variable intercepts under the standard parameterization as the population value is zero.
    \end{tablenotes}
  \end{threeparttable}
}
\end{table}
Across all sample sizes and conditions none of the MIIV solutions exhibited nonconvergence. For the WLSMV estimator convergence problems arose in the smaller sample sizes. Across the two parameterizations approximately $10\%$ and $0.5\%$ of the WLSMV solutions failed to converge for the sample sizes of $100$ and $200$, respectively. Nonconvergence of WLSMV was more frequent under the alternative parameterization. Both estimators exhibited a large portion of nonpositive definite covariance matrices at the smaller sample sizes. Here the percentage of nonpositive definite solutions was similar across estimators and ranged from approximately $50\%$ at a sample size of $100$ to $25\%$ at a sample size of $200$. The rate of nonpositive definite matrices decreased considerably in larger sample sizes.  Details on the frequency of nonconverged and nonpositive definite solutions are provided in Table~\ref{baddata}.

\subsubsection{Point Estimates}
The percentage of relative bias for point estimates and standard errors are provided in  Table~3. Here any percentage of relative bias exceeding $5\%$ in absolute value is presented in bold typeface.  First we consider the summary statistics for the sample of datasets that does not include solutions with nonpositive definite matrices.  Under the standard parameterization the MIIV estimator exhibited acceptable bias in point estimates for nearly all parameter sets and sample sizes.  The only exceptions being the ordinal variable factor loadings and error variances at the smallest sample size of $100$.  Under the alternative parameterizations the intercepts and factor variances showed nontrivial bias at the two smallest sample sizes, $N=100,200$. The factor loadings and error variances for the ordinal variables also exhibited a slight negative bias at $N=100$ under the alternative parameterization.\\

For the WLSMV estimator the pattern of relative bias is more nuanced.  Under the standard parameterization, relative bias for WLSMV exceeds the $5\%$ cutoff for all parameters except for the thresholds at $N=100$.  Additionally, the continuous variable factor loadings, regression coefficients and factor variances also exhibited nontrivial bias at $N=200$. Under the alternative paramaterization bias of the WLSMV estimates were nontrivial for both the variance parameters at the two smallest sample sizes and  the factor loadings and regression coefficients at the smallest sample size only. The pattern of problematic bias for the WLSMV was similar under the alternative parameterization except for the error variances, which no longer showed meaningful bias.

Next we consider the situation in which all converged datasets are included in the analysis. The rationale for including solutions with improper solutions is based on the idea that (1) this situation may be quite common in practical modeling situations and (2) parameter estimates from these solutions are required to obtain unbiased summaries of the estimator's sampling variability.  For the MIIV estimator, a larger amount of bias in the point estimates was observed when all converged datasets were included in the analysis. This bias was nontrivial at the smallest sample size of $100$ for the factor loadings and variance parameters under the standard parameterization, and the factor loadings, regression coefficients and error variances under the alternative parameterization.  For the WLSMV estimator the factor variance parameters exhibited problematic levels of bias at sample sizes of $100$ and $200$, as well as the error variances at the smallest sample size of $100$ under the standard parameterization only. 

\subsubsection{Standard Errors}

Again we begin by examining the sample of datasets where solutions with nonpositive definite matrices have been excluded.  For the MIIV estimator the bias of the standard errors speaks to both the performance of the estimator and the accuracy of the analytic results derived earlier. Under the standard parametrization standard error bias exceeded nominal levels for the thresholds and variance parameters at the smallest sample size, and for the error variances at a sample size of $200$.Under the alternative parameterizations only the thresholds at the smallest sample size exhibited problematic bias. Compared to the MIIV estimator WLSMV exhibited a greater degree of standard error bias at smaller sample sizes. This finding was consistent across parameter type and this bias was problematic for the majority of parameters at a sample size of $100$. 

Including all converged datasets in the analysis increased the standard error bias across both estimators.  For the MIIV estimator, the regression coefficients and variance parameters exhibited problematic bias at the two smallest sample sizes across both parameterizations. Under the alternative parameterization the MIIV standard errors for the thresholds, intercepts and factor loadings also showed nontrivial biases at the two smallest sample sizes.  The factor variance standard errors were the only MIIV estimates across all simulation conditions to show appreciable bias at a sample size of $400$ or larger.  When including all converged datasets the WLSMV estimator also showed an increase in SE bias across the two parameterizations.  Under the standard and alternative parameterizations the factor loadings, regression coefficients, and variance parameters showed problematic bias at sample sizes of $100$ and $200$. Under the alternative parameterization only, the WLSMV intercepts and thresholds also showed problematic bias at the smaller sample sizes. For the WLSMV estimator the only parameters that exhibited problematic bias at sample sizes of $400$ or greater were the regression coefficient and factor variance standard errors.

\begin{figure}[h!]
  \includegraphics[width=1\textwidth]{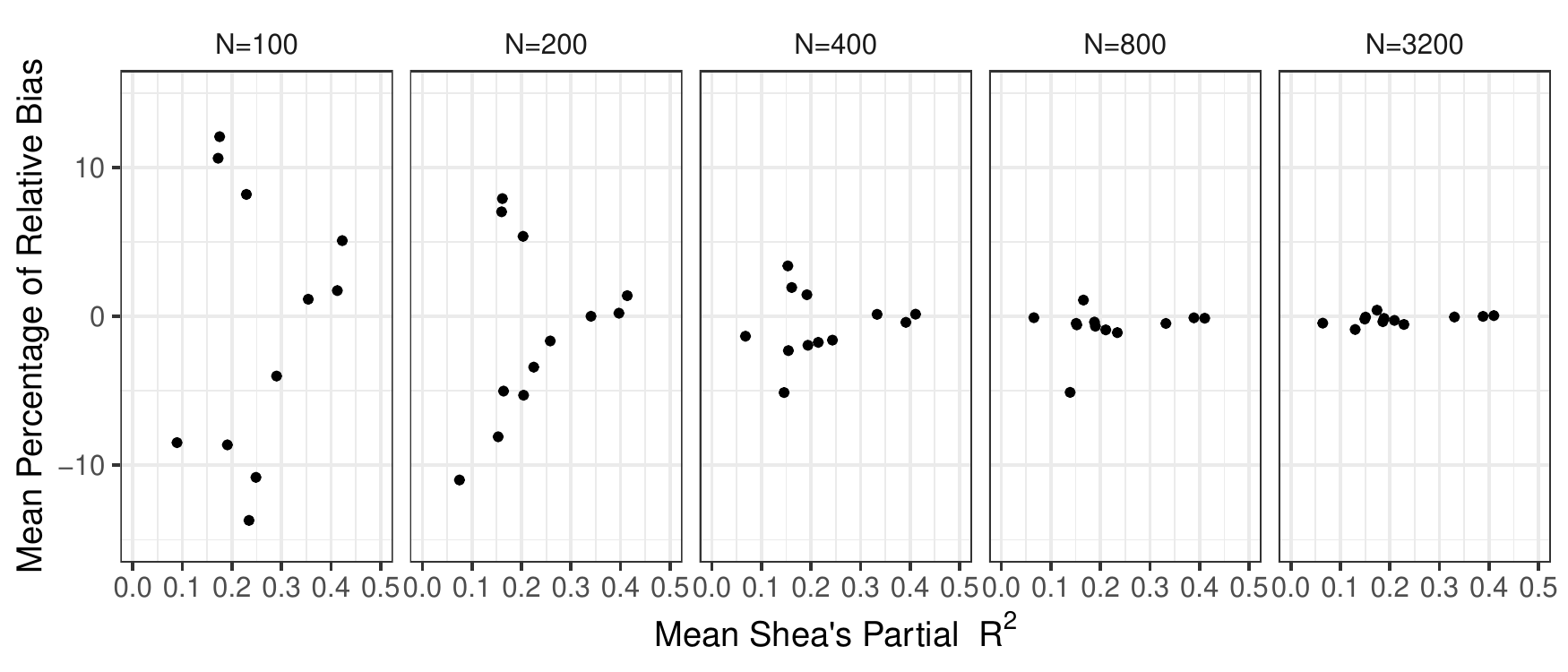}
  \caption{Instrument Strength and Relative Bias by Parameter and Sample Size}
  \label{weak}
\end{figure}

\subsubsection{Instrument Strength}

\textcolor{black}{Equation-level instrument strength varied across all replications included in our simulations. For this reason it is possible to examine the impact of instrument strength on outcome measures such as relative bias.  In Figure~\ref{weak} instrument strength as indexed by Shea's partial $R^2$ is aggregated within each parameter and presented across sample sizes. From this illustration it is clear that smaller $R^2_S$ values are associated with more variability in bias, and this is most clear at the smaller sample sizes. The variability in the percentage of relative bias also tends to decrease with increases in $R^2_S$ at each sample size, though this is less noticeable at the largest sample sizes. Unfortunately, as the sampling distribution of $R^2_S$ has not been derived it is not possible to examine the impact of instrument strength on inferential test size for our simulations.}

\section{Discussion}

\textcolor{black}{\citet{jin2018} recently reviewed the evidence comparing the PIV estimator to DWLS and ULS for factor analysis models and concluded the PIV approach produces estimates as accurate as ULS and DWLS if the model is correctly specified and more robust estimates if the model is misspecified. We agree, but note that the PIV has several restrictions which if removed would make the PIV an even more powerful alternative.  Among these is that \citet{bollen2007} developed the PIV estimator for only binary and ordinal endogenous variables and not a mixture of categorical and continuous variables.  Also, they and the researchers who followed them, used numerical derivatives to develop the asymptotic covariance matrix of the parameter estimates rather than using analytic derivatives.  Analytic derivatives are valuable because they can increase both the efficiency of the estimation and the accuracy of estimates.   In addition, with the analytic derivatives in hand, the PIV estimator can be applied to any means and covariance matrix as input along with their asymptotic covariance matrix.  In addition, nearly all of the work on polychoric correlations has focused on correlations rather than means, variances, and covariances.   An additional property that would be desirable for both WLSMV and PIV would be the ability to input the means, variances, and covariances of the underlying variables.  Researchers using traditional SEM with continuous endogenous variables can estimate models based on the covariance matrix and means.  Using these as input facilitates comparisons of effects both over time and across different groups.}

\textcolor{black}{Our paper overcomes all of these restrictions.  We develop the PIV estimator to permit a mixture of binary, ordinal, or continuous endogenous variables.  We use differential matrix calculus to elaborate the analytic derivatives for a general set of patterned matrices that permit us to use the multivariate delta method to derive the asymptotic standard errors of the parameter estimates.  In addition, we pick up a thread originally elaborated for SEM by \citet{joreskog2002} regarding the construction of ordinal variable indicators. To our knowledge we are the first to realize and detail the full generality of this idea.  \citet{joreskog2002} originally considered the estimation of mean and variance parameters for the underlying variable of an ordinal variable whose first two thresholds are constrained to zero and one, respectively.}

\textcolor{black}{The choice of using the first two thresholds and the fixed constants of zero and one is often a convenient choice as it sets a unit scale for the threshold values, however, as demonstrated in the empirical example, a number of other possibilities are both useful and possible. Our general method enables the estimation of means, variances, and covariances for the underlying variables of ordinal observed variables with three or more categories along with continuous variables.  By so doing, we create means and covariance matrix as is common when researchers analyze all continuous variables.  As part of this, we used methods that allow analyst to input known threshold values.  This permits researchers to take advantage of the cutoff points (e.g., 8 years of education, less than high school, etc.) for ordinal variables that are collapsed versions of continuous variables.  By providing the algebraic transformations required for obtaining these parameterizations we hope to contribute to additional work on ordinal variable parameterizations, including multiple group models where measurement invariance testing often requires complicated parameter constraints. Furthermore, our matrix derivatives take this scaling flexibility into account.  We note that analysts can use this method of generating means, variances, and covariances with continuous and ordinal variables with DWLS and ULS methods as well as PIV.}

\textcolor{black}{In a simulation study we have shown the performance of the proposed point estimator and standard errors to be equivalent to the popular WSLMV estimator under a number of sample sizes and correct model parameterizations. Importantly, we have only considered correctly specified models in our simulation study. In doing so we have ignored the scenarios most likely encountered in applied research (model misspecification) and also the conditions under which the MIIV estimator is most likely to demonstrate superior performance over the system-wide estimators.  Our empirical example using education in a continuous and ordinal form allowed us to enter fixed thresholds (e.g., 8 years, 12 years) as well as to estimate the thresholds that correspond to other ordinal categories where the years of education are uncertain.}
  
\textcolor{black}{Finally, the proposed developments not only support the current goal of accommodating mixed data types but also support future extensions of the MIIV framework. Other researchers can use the analytic derivatives we present in a number of important situations not yet considered in the MIIV framework, such as the handling of missing data in a manner similar to that proposed by \citet{yuan2000} and \citet{savalei2014} and the development of accurate standard errors for data with complex dependencies such as time-series data. For this reason, and the extensions to multiple-group modeling discussed earlier, we view the developments made here as an important building block for future development of the PIV estimator.}

\newpage

\section{Appendix A}

\section{Notation and Algebraic Results}
\subsubsection{Vec and related operators.}
Our derivations make use of the $\mathrm{vec}$ and related operators for transforming a matrix into a vector.  Although these operators are commonly encountered in multivariate analysis there representations vary considerably by author.  For this reason we will provide definitions corresponding to our own usage. For a $p \times q$ matrix, $\mathbf{X}$, the $\mathrm{vec}$ operator is used to stack columnwise the $q$ columns of $\mathbf{X}$ into a $pq \times 1$ vector without regard for any repeated or constant elements. Consider the matrix $\mathbf{A}_{p,q}$, where $\mathbf{a}_{1},\dots, \mathbf{a}_{q}$ are the columns of $\mathbf{A}_{p,q}$ taken in lexicon order then $\mathrm{vec}\:\mathbf{A}_{p,q} = [\mathbf{a}_{1} , \dots,\mathbf{a}_{q} ]^{'}$. \\
The $\upsilon(\cdot)$ operator is also used heavily in these derivations. Here, $\upsilon(\cdot)$, can be understood as a generalization of the $\mathrm{vech}(\cdot)$ (\emph{vector-half}) operator for symmetric matrices, or the $\mathrm{vecp}(\cdot)$ operator for strictly lower-triangular matrices, to any patterned matrix.  A $p \times q$ matrix is labeled \emph{patterned} if it contains $p^{*}=pq-s-v$ mathematically independent and variable elements, where $s$ and $v$ are the number of repeated, and constant elements, respectively. Covariance and correlation matrices are two examples of patterned matrices. For a $p \times p$ covariance matrix, $\mathbf{S}$, $p^{*} =  1/2p(p+1)$, $s =  1/2p(p+1)$, and $v=0$. Likewise, for the $p \times p$ correlation matrix, $\mathbf{P}$, $p^{*} =  1/2p(p-1)$, $s =  1/2p(p-1)$, and $v=p$.  Consider the covariance matrix,  $\mathbf{S}_{2,2}$, then $\upsilon(\mathbf{S})=[{s}_{1,1}, {s}_{2,1}, {s}_{2,2}]^{'}$. Similarly, for the correlation matrix, $\mathbf{P}_{2,2}$, $\upsilon(\mathbf{P})=[{s}_{2,1}]^{'}$. Generally, for any patterned matrix $\mathbf{X}_{p,q}$, $\upsilon(\mathbf{X})$ will be a $p^{*} \times 1$ vector.
\subsubsection{The Kronecker product.}
For the matrices $\mathbf{X}_{p,q}$ and $\mathbf{A}_{r,s}$, we define the Kronecker product as $\mathbf{X} \otimes \mathbf{B} = (x_{i,j}\mathbf{A})_{pr \times qs}$, for $i =1,\dots,p$ and $j =1,\dots,q$. A useful result linking the Kronecker product to the $\mathrm{vec}$ operator states $\mathrm{vec}(\mathbf{A}_{m,n}\mathbf{B}_{n,q}\mathbf{C}_{q,r}) = (\mathbf{C}^{'} \otimes \mathbf{A})\mathrm{vec}\mathbf{B}$.  Other useful properties of the Kronecker product utilized in these derivations for simplifying resultant expressions are $(\mathbf{A} \otimes \mathbf{B}) (\mathbf{C} \otimes \mathbf{D}) = \mathbf{AC} \otimes \mathbf{BD}$ and  $(\mathbf{A} \otimes \mathbf{B})^{'} = \mathbf{A}^{'} \otimes \mathbf{B}^{'}$. 
\subsubsection{The commutation matrix.} Commutation (or vec-permutation) matrices can be used to translate between vectors $\mathrm{vec}\:\mathbf{X}$ and $\mathrm{vec}\:\mathbf{X}^{'}$.  The vec-permutation operator, $\mathbf{K}_{p,q}$, is defined such that $\mathrm{vec}\:\mathbf{X}_{p,q} = \mathbf{K}_{p,q} \mathrm{vec}\:\mathbf{X}^{'}$. The commutation matrix plays a central role in the formulation of matrix derivatives using the $vec$ operator and the following derivative is used throughout. Consider the $m \times n$ matrix $\mathbf{X}$, then  $\partial\mathrm{vec}(\mathbf{X}^{'}) / \partial\mathrm{vec}(\mathbf{X})^{'} = \mathbf{K}_{m,n}$. Note also that $\partial\mathrm{vec}(\mathbf{X}) / \partial\mathrm{vec}(\mathbf{X})^{'}  = \mathbf{I}_{mn}$.
\subsubsection{Matrix Derivatives and L-Structured Matrices}  The results herein require taking partial derivatives with respect to \emph{lower-triangular}, \emph{strictly lower-triangular}, \emph{diagonal} and \emph{arbitrarily patterned} matrices. Furthermore, the solution matrices resulting from these matrix derivatives are themselves often known apriori to be symmetric or patterned. For these reasons we rely on a number of results detailed by \citet{magnus1983} and \citet{magnus1986}  for L-structured matrices. The use of L-structures allows us to derive our results in the most general way possible across the different parameterizations available.  The following properties of L-Structured matrices are used throughout, $\mathrm{vec}(\mathbf{X})=\boldsymbol{\Delta}\upsilon(\mathbf{X})$ and $\upsilon(\mathbf{X})=\boldsymbol{\Delta}^{+}\mathrm{vec}(\mathbf{X})$. \\

It is useful to consider $\boldsymbol{\Delta}$ as a generalized duplication matrix, and $\boldsymbol{\Delta}^{+}$ as a generalized elimination matrix.  If $\mathbf{X}$ is a \emph{symmetric} $\boldsymbol{\Delta}$ is $p^2 \times p(p+1)/2$, while $\boldsymbol{\Delta}$ is $p^2 \times p(p-1)/2$ if $\mathbf{X}$ is \emph{strictly lower-triangular}. The most interesting case occurs when $\mathbf{X}$ exhibits an arbitrary constellation of free, fixed and repeating elements.  In this case a general method is needed for constructing $\boldsymbol{\Delta}$ and $\boldsymbol{\Delta}^{+}$ when the specific patterning of $\mathbf{X}$ is unknown (prior to the analysis). Fortunately, a result for this specific case was derived by \citet[Definition 6.1.1]{nel1980}. In the case of arbitrary patterning $\boldsymbol{\Delta}$ is $p^2 \times p^{*}$.

Extending these properties to the case of matrix derivatives it can be shown that
$\frac{\partial \mathrm{vec} (\mathbf{X})}{\partial\: \upsilon(\mathbf{X})^{'}} = \boldsymbol{\Delta}$ and $
\frac{\partial \:\upsilon (\mathbf{X})}{\partial\: \mathrm{vec} (\mathbf{X})^{'}} = \boldsymbol{\Delta}^{+}$. It follows that if the matrix function $\mathbf{Z} = f(\mathbf{X})$,
\begin{IEEEeqnarray}{rClrCl} 
\label{patderiv3}
\frac{\partial \mathrm{vec} (\mathbf{Z})}{\partial \upsilon(\mathbf{X})^{'}} & = & 
\frac{\partial \mathrm{vec} (\mathbf{Z})}{\partial \mathrm{vec}(\mathbf{X})^{'}}
\frac{\partial \mathrm{vec} (\mathbf{X})}{\partial \upsilon(\mathbf{X})^{'}} & = & 
\frac{\partial \mathrm{vec} (\mathbf{Z})}{\partial \mathrm{vec}(\mathbf{X})^{'}}
 \boldsymbol{\Delta},
\end{IEEEeqnarray}
\noindent and if $\mathbf{Z}$ is also patterned,
\begin{IEEEeqnarray}{rClrCl} 
\label{patderiv3}
\frac{\partial \upsilon (\mathbf{Z})}{\partial\: \upsilon(\mathbf{X})^{'}} & = & 
\frac{\partial \upsilon (\mathbf{Z})}{\partial \mathrm{vec}(\mathbf{Z})^{'}}
\frac{\partial \mathrm{vec} (\mathbf{Z})}{\partial \mathrm{vec}(\mathbf{X})^{'}}
\frac{\partial \mathrm{vec} (\mathbf{X})}{\partial \upsilon(\mathbf{X})^{'}} & = & 
 \boldsymbol{\Delta}^{+}
\frac{\partial \mathrm{vec} (\mathbf{Z})}{\partial \mathrm{vec}(\mathbf{X})^{'}}
 \boldsymbol{\Delta}.
\end{IEEEeqnarray}

\subsubsection{Derivatives of Common Matrix Functions}
The following derivatives are used throughout and will be restated here for clarity. For the following results suppose $\mathbf{X}$ is $m \times n$, $\mathbf{U}$ is $p \times q$, and $\mathbf{V}$ is $q \times r$, where both  $\mathbf{U}$ and  $\mathbf{V}$ are matrix functions of $\mathbf{X}$, then
\begin{IEEEeqnarray}{rCl} 
\label{sumrule}
\frac{\partial \:\mathrm{vec}\: (\mathbf{U+V})}{\partial\: \mathrm{vec}\: (\mathbf{X})^{'}} & = &
\frac{\partial \:\mathrm{vec}\: (\mathbf{U})}{\partial\: \mathrm{vec}\: (\mathbf{X})^{'}} +
\frac{\partial \:\mathrm{vec}\: (\mathbf{V})}{\partial\: \mathrm{vec}\: (\mathbf{X})^{'}},
\end{IEEEeqnarray}
\noindent and
\begin{IEEEeqnarray}{rCl} 
\label{productrule}
\frac{\partial \:\mathrm{vec}\: (\mathbf{UV})}{\partial\: \mathrm{vec}\: (\mathbf{X})^{'}} & = &
(\mathbf{V} \otimes \mathbf{I}_{p})^{'}
\frac{\partial \:\mathrm{vec}\: (\mathbf{U})}{\partial\: \mathrm{vec}\: (\mathbf{X})^{'}} +
(\mathbf{I}_{r} \otimes  \mathbf{U} )
\frac{\partial \:\mathrm{vec}\: (\mathbf{V})}{\partial\: \mathrm{vec}\: (\mathbf{X})^{'}}.
\end{IEEEeqnarray}
\noindent In addition we state a general rule for taking derivatives of matrix inverses, specifically if $\mathbf{Y} = \mathbf{X}^{-1}$, then
\begin{IEEEeqnarray}{rCl} 
\label{invrule}
\frac{\partial \:\mathrm{vec} (\mathbf{Y})}{\partial\: \mathrm{vec} (\mathbf{X})^{'}} & = &
-(\mathbf{X}^{-1'} \otimes \mathbf{X}^{-1}).
\end{IEEEeqnarray}
\noindent Now suppose $\mathbf{X}$ is a symmetric matrix, $\mathbf{A}$ is a matrix of constants, and $\mathbf{Y} = \mathbf{X}\mathbf{A}\mathbf{X}$ using successive applications of \eqref{productrule} we can show that, 
\begin{IEEEeqnarray}{rCl} 
\label{ranresult3}
\frac{\partial \:\mathrm{vec}(\mathbf{Y})}{\partial\: \mathrm{vec}(\mathbf{X})^{'}} 
&=&
(\mathbf{AX} \otimes \mathbf{I})^{'} +
(\mathbf{I} \otimes  \mathbf{XA} ). 
\end{IEEEeqnarray}
Suppose instead that $\mathbf{A}$  and $\mathbf{B}$ are matrices containing constant elements and $\mathbf{Y} = \mathbf{A}^{'}\mathbf{X}^{-1}\mathbf{B}$, then
\begin{IEEEeqnarray}{rCl} 
\label{matderiv2}
\frac{\partial \:\mathrm{vec}(\mathbf{Y})}{\partial\: \mathrm{vec}(\mathbf{X})^{'}} 
&=&
-(\mathbf{B}^{'}\mathbf{X}^{-1'} \otimes \mathbf{A}^{'}\mathbf{X}^{-1}).
\end{IEEEeqnarray}

\bibliographystyle{apacite}
\bibliography{example}
\end{document}